\documentclass[preprint,nofootinbib,aps,superscriptaddress,eqsecnum]{revtex4-1}
 \pdfoutput=1
\usepackage{amsmath,amssymb,graphicx,subfigure}

\newcommand{\beq}{\begin{equation}}
\newcommand{\eeq}{\end{equation}}
\newcommand{\bea}{\begin{eqnarray}}
\newcommand{\eea}{\end{eqnarray}}
\newcommand{\Z}{\mathbb{Z}}

\def\nn{\nonumber}

\def\ra{\rightarrow}
\def\mpl{M_{\rm Pl}}

\newcommand{\MS}{\overline{\mbox{\sc ms}}}

\def\ie{ {\em i.e.,\ }}

\def\mpl{M_{\rm Pl}}

\begin{document}
\title{Exploring the Hyperchargeless Higgs Triplet Model up to the Planck Scale }
 \author{Najimuddin Khan}\email{phd11125102@iiti.ac.in, khanphysics.123@gmail.com} 
\affiliation{Discipline of Physics, Indian Institute of Technology Indore,\\
 Khandwa Road, Simrol, Indore - 453~552, India \vspace{1.80cm}}


\begin{abstract}\vspace*{10pt}
We examine an extension of the SM Higgs sector by a Higgs triplet taking into consideration the Higgs-like particle discovery at the LHC with mass around 125 GeV. We evaluate the bounds on the scalar potential through the unitarity of the scattering-matrix. Considering with and without $\Z_2$-symmetry on the extra triplet,
we derive constraints on the parameter space. We identify the region of the parameter space that corresponds to the stability and metastability of the electroweak vacuum. We also show that at large field values the scalar potential of this model is suitable to explain inflation.
\end{abstract}

\maketitle

\section{Introduction}
The revelation of the Higgs boson~\cite{Aad:2012tfa,Chatrchyan:2012ufa,Giardino:2013bma} in 2012 at the Large Hadron Collider (LHC), confirmed the existence of all the Standard Model (SM) particles and the Higgs mechanism to be responsible for electroweak symmetry breaking (EWSB).
So far, the LHC, operated with $pp$ collision energy at $\sqrt{s}\sim$ 8 and 13 TeV has not found any signature of new physics beyond the standard model (BSM).
However, various theoretical issues such as the hierarchy problem related to the mass of the Higgs, mass hierarchy and mixing patterns in the leptonic and quark sectors suggest the need for new physics beyond the SM. Different experimental observations such as non-zero neutrino mass, baryon-antibaryon asymmetry in the Universe, mysterious nature of dark matter (DM) and dark energy, inflation in the early Universe indicate the existence of new physics.
Moreover, the measured properties of the Higgs boson with mass $\sim$125 GeV are consistent with those of the scalar doublet as predicted by the SM. However, the experimental data~\cite{Khachatryan:2016vau} still comfortably allow an extended scalar sector, which may also be responsible for the EWSB.

The present experimental values of the SM parameter of the Lagrangian indicate that if the validity of the SM is extended up to the Planck mass $(\mpl =1.2\times10^{19}~{\rm GeV})$, a second, deeper minimum is located near the Planck mass such that the EW vacuum is metastable.
The transition lifetime of the EW vacuum to the deeper minimum is finite $\tau_{EW} \sim 10^{300}$ years~\cite{Degrassi:2012ry,Buttazzo:2013uya,
Khan:2014kba,Khan:2015ipa,Masina:2012tz,EliasMiro:2011aa, Branchina:2013jra, Branchina:2014usa, Branchina:2014rva, Branchina:2016bws, Branchina:2015nda,Bentivegna:2017qry}. The EW vacuum remains metastable even after adding extra scalar particles to the SM which have been discussed in Refs.~\cite{Khan:2014kba,Khan:2015ipa,Datta:2016nfz,Basso:2013nza,Fischer:2016rsh}.

In this work, we add a real hypercharge $Y=0$ scalar triplet to the SM.
In the literature, this model is termed as the hyperchargeless Higgs
triplet model, HTM ($Y=0$)~\cite{Blank:1997qa}.
We consider both the neutral $CP$-even component of the SM doublet and the extra scalar triplet take part in the EWSB.
Including radiative corrections, we check the validity of the parameters of the model up to the Planck mass $\mpl$.
We review various theoretical and experimental bounds of this model. In this work, we especially discuss the unitary bounds of the quartic couplings of the scalar potential. To the best of our
knowledge, the unitary bounds of this model were not discussed in the literature.
Next, we impose a $\Z_2$-symmetry such that an $odd$ number of scalar particles of the triplet do not couple with the SM particles.
The lightest neutral scalar particle does not decay and becomes stable.
This scalar field can be taken as a viable DM candidate which may fulfill the relic abundance of the Universe.
In this context, it is instructive to explore whether these extra scalars can also prolong the lifetime of the Universe.
In this model,
we find new regions in the parameter space of this model in which the EW vacuum remains metastable.
We also consider that the extra neutral scalar field (also compatible as a viable dark matter candidate) can act as an inflaton. We show that this scalar field is able to explain the inflationary observables.

A detailed study of the HTM ($Y=0$) parameter space which is valid up to 1 TeV, has been performed in Refs.\cite{Forshaw:2003kh}.
Two different renormalization schemes, electroweak precision, and decoupling of Higgs triplet scenario have been discussed in Ref.\cite{Chen:2008jg}.
Using the electroweak precision test (EWPT) data and one-loop correction to the $\rho$ parameter, the Higgs mass range has been predicted in Refs.\cite{Chen:2006pb,Chen:2005jx, Chankowski:2006hs,Forshaw:2001xq,Khandker:2012zu}.
The detailed structure of the vacuum of the scalar potential at the tree-level has been studied in the Ref.\cite{FileviezPerez:2008bj}.
The constraints on the parameter spaces from the recent LHC $\mu_{\gamma\gamma}$ and $\mu_{Z\gamma}$ data have been discussed in Ref.~\cite{Wang:2013jba}.
The LHC and future collider experiments with high luminosity can be used as an useful tool to detect these extra scalar particles through vector bosons scatterings~\cite{Khan:2016mdq}.
More recently, the inert scalar triplet has been investigated in the context of dark matter direct and indirect detection~\cite{Araki:2011hm,Ayazi:2014tha, Ayazi:2015mva}.
The heavier inert fields can decay through one-loop via extra Majorana fermions~\cite{JosseMichaux:2012wj,Lu:2016ucn}.
This model has the required ingredients to realize a successful leptogenesis which can explain the matter asymmetry in the Universe~\cite{JosseMichaux:2012wj,Lu:2016ucn}.
The multi-component dark matter have been investigated~\cite{Fischer:2011zz,Fischer:2013hwa} in HTM with extra scalar multiplets of $SU(2)$ representation.

The paper is organized as follows. Section~\ref{model} starts with a detailed descriptions of HTM ($Y=0$) model. We discuss detailed constraints in Sec.~\ref{sec:constraints}. Considering the lightest $\Z_2$-odd neutral particle as a viable DM, we analyze the scalar potential up to the Planck mass and identify regions of parameter space corresponding to the stable and metastable EW vacuum in Sec.~\ref{sec:ITM}. We explain inflation as well in Sec.~\ref{sec:infl}. Finally we
conclude in Sec.~\ref{sec:conclu}.

\section{Model}\label{model}
We consider a model with a real Higgs doublet, $\Phi$, and a real, isospin $I=1$, hypercharge $Y=0$ triplet $T$. The extra scalar triplet consists of a pair of singly-charged fields and a $CP$-even neutral scalar field.
The doublet and triplet scalar are conventionally written as~\cite{Chen:2008jg}
\begin{eqnarray}
\Phi=\left(
\begin{array}{c}
{G_{1}^+} \\
\frac{1}{\sqrt{2}}(v_1+{h^0}+i{G^0})
\end{array}
\right)\, ,
\qquad \qquad
T=\left(
\begin{array}{c}
{\eta^+}\\
v_2+{\eta^0}\\
-{\eta^-}
\end{array}
\right)\, .\label{Tripfield1}
\end{eqnarray}

The kinetic part of the Lagrangian is given by
\begin{equation}
{\cal L}_k=\mid D_\mu \Phi \mid^2+\frac{1}{2}\mid D_\mu T\mid^2 \, ,
\end{equation}
where the covariant derivatives are defined as,
\beq
D_\mu \Phi = \biggl( \partial_\mu +i \frac{g_2}{ 2}
 \sigma^aW^a + i\frac{g_1}{ 2} Y B_\mu\biggr)\Phi
~~~\text{and}~~~
D_\mu T = \biggl(\partial_\mu +i g_2 t_aW^a\biggr)T\, ,
\eeq
where, $W^{a}_\mu$ ($a$=1,2,3) are the ${ SU(2)_L}$ gauge bosons, corresponding to three generators of ${ SU(2)_L}$ group and $B_\mu$ is the ${ U(1)}_Y$ gauge boson.
$\sigma^a ~(a=1,2,3)$ are the Pauli matrices,
and $t_a$ can be written as follows
\begin{equation}
t_1=\frac{1}{ \sqrt{2}}
\left(
\begin{array}{ccc}
0&1&0\\
1&0&1\\
0&1&0
\end{array}\right),\quad
t_2=\frac{1}{\sqrt{2}}
\left(\begin{array}{ccc}
0&-i&0\\
i&0&-i\\
0&i&0
\end{array}\right),\quad
t_3=\left(\begin{array}{ccc}
1&0&0\\
0&0&0\\
0&0&-1
\end{array}\right)\, .
\end{equation}

The scalar potential is such that both the neutral $CP$-even component of the SM doublet and the extra scalar triplet receive vacuum expectation values (VEVs), and thus take part in the EWSB.
After EWSB, one of the linear combinations of charged scalar fields of scalar doublet and the triplet is eaten by the $W$ boson which becomes massive, other orthogonal combinations of these fields become massive charged scalar fields. Similarly, a pseudoscalar of scalar doublet become the longitudinal part of massive $Z$ gauge boson. This scalar may give rise to a signature through the scattering of vector bosons~\cite{Khan:2016mdq} in collider experiments. The spontaneous EWSB generates masses for the $W$ and $Z$ bosons as
\begin{eqnarray}
M_W^2 &=& \frac{g_2^2}{ 4}\biggl( v_1^2+4 v_2^2\biggr),
\nonumber ~~ \text{and}~~
M_Z^2 = \frac{g_2^2}{ 4 c_\theta^2}v_1^2\, ,
\end{eqnarray}
where, $c_W\equiv \cos\theta_W=g_2/\sqrt{g_1^2+g_2^2}$ and $s_W\equiv\sin\theta_W$.
The scalar doublet VEV $v_1$ and the triplet VEV $v_2$ are related to the SM VEV by $v_{SM}(\equiv246.221~\text{GeV})=\sqrt{v_1^2+4 v_2^{ 2}}$.

One can see that this  model violates custodial symmetry at tree level
\begin{eqnarray}
\rho&=&\frac{M_W^2}{M_Z^2 c_W^2}
=
1+4\frac{v_2^{2}}{v_1^2}
\, .
\end{eqnarray}
The experimental value of $\rho$ is $1.0004\pm 0.00024$~\cite{Agashe:2014kda} at 1$\sigma$. Hence, $\delta \rho\approx0.0004\pm0.00024$ and we will adopt the bound $\delta \rho\leq 0.001$. This puts a stringent constraints on  $v_2$ and we get $v_2$ should be less than 4 GeV.

The tree-level scalar potential with the Higgs doublet and
the real scalar triplet is invariant under $SU(2)_L\times U(1)_Y$ transformation. This is given by
\begin{eqnarray}
V(\Phi,T)&=&\mu_1^2\mid \Phi\mid^2
+\frac{\mu_2^2}{ 2}\mid T\mid^2
+{\lambda_1} \mid \Phi\mid^4 
+\frac{\lambda_2}{ 4}\mid T\mid^4
\nonumber \\ &&
+\frac{\lambda_3}{ 2}\mid \Phi\mid^2\mid T \mid^2
+\lambda_4  \Phi^\dagger \sigma^a \Phi T_a \, .
\label{ScalarpotTrip}
\end{eqnarray}
We have the following minimization conditions of the tree-level scalar potential
\bea
\mu_1^2 &=&\frac{1}{2} \lbrace 2 \lambda_4 v_2 -  (2 \lambda_1 v_1^2 +\lambda_3 v_2^2 )\rbrace ,\\
\label{min1}
\mu_2^2 &=& \frac{1}{2 v_2}\lbrace \lambda_4 v_1^2 - \lambda_3 v_1^2 v_2 - 2 \lambda_2 v_2^3\rbrace .
\label{min2}
\eea
After electroweak symmetry breaking, the squared mass matrix can be expressed as $6\times 6$ for the scalar fields ($G^\pm_1,\eta^\pm,\eta^0$ and $h^0$). This matrix is composed of three $2\times 2$ submatrices with bases, ($G_1^+$, $\eta^+$), ($G_1^-$, $\eta^-$) and ($h^0$, $\eta^{0}$). After rotating these fields into the mass basis, we get four physical mass eigenstates ($H^\pm,h,H$). The remaining two states ($G^\pm$) and $G^0$ become the massless Goldstone bosons.

The physical masses of the particles are given by
\bea
M^2_{h} &=& \frac{1}{2}\left[(B+A)-\sqrt{(B-A)^{2}+4 C^{2}}\right],\nn\\
M^2_{H} &=& \frac{1}{2}\left[(B+A)+\sqrt{(B-A)^{2}+4 C^{2}}\right],\label{htmy0H0}\\
M_{H^\pm}^2 &=&\lambda_4 \frac{(v_1^2 + 4 v_2^2)}{2 v_2}\nn,
\eea
where,
\beq
A = {2 \lambda}_1 v_1^2,
~~~
B =\frac{{\lambda}_4 v_1^2 + 4 {\lambda}_2 v^{3}_2}{2 v_2},~~~\text{and}~~~C =-{\lambda}_4 v_1 + {\lambda}_3 v_1 v_2.\label{ctilde}
\eeq

The mixing between the doublet and triplet in the charged and $CP$-even scalar sectors are respectively given by
\begin{eqnarray}
\left(
\begin{array}{c}
h\\
H
\end{array}
\right)
&=&
\left(
\begin{array}{cc} 
c_\gamma & s_\gamma \\
-s_\gamma & c_\gamma 
\end{array}
\right)
\left(
\begin{array}{c}
{h^0}\\
{\eta^0}
\end{array}
\right)\, ,\\
\left(
\begin{array}{c}
G^\pm\\
H^\pm
\end{array}
\right)
&=&
\left(
\begin{array}{cc} 
c_\beta & s_\beta \\
-s_\beta & c_\beta 
\end{array}
\right)
\left(
\begin{array}{c}
{G_{1}^\pm}\\
{\eta^\pm}
\end{array}
\right)\, ,
\end{eqnarray}
where,
\beq
s_\gamma(\equiv\sin\gamma) = \sqrt{\frac{\sqrt{(B-A )^2 + 4C^2}-(B - A)}{2 \sqrt{(B-A)^2 + 4C^2}}} ~~~\text{and}~~~\tan\beta = \frac{2 v_2}{v_1}.\nn 
\eeq

In large $\mu_2^2$ and small $v_2$ limit, one can express $\sin\gamma$ and $\sin\beta$ as
\beq
s_\gamma ~= \sqrt{\frac{1}{2}-\frac{1}{2\sqrt{1+ 16\frac{v_2^2}{v_1^2}}}}~\approx~0 ~~~\text{and}~~~
s_\beta ~= \frac{2 v_2}{\sqrt{v_1^2 + 4 v_2^2}}~\approx~0.\nn
\eeq 

In these limits, the quartic $\lambda_{1,2,3}$ and $\lambda_4$ can be written as
\beq
\lambda_1=\frac{M_h^2}{2 v_1^2},~~
\lambda_2=\frac{2 (M_{H}^2 -M_{H^\pm}^2)}{  v_1^2 s_\beta^2},~~
\lambda_3=\frac{2( M_{H^\pm}^2- (s_\gamma/s_\beta) M_{H}^2)}{ v_1^2},~~
\lambda_4=\frac{s_\beta M_{H^\pm}^2}{v_1}.
\label{AppMass}
\eeq
In the same limit, if $M_{H^\pm}$ and $M_H$ are very heavy compared to $M_h$, then $M_{H^\pm}$ and $M_H$ become degenerate (see eqns.~\ref{htmy0H0} and \ref{ctilde}).
If the mass difference between $M_{H^\pm}$ and $M_H$ is large, then the quartic couplings $\lambda_{2,3}$ will violate the perturbativity and unitarity (see subsections~\ref{ssec:perturbativity} and~\ref{ssec:unitarity}) bounds.

The SM gauge symmetry, $SU(2)_L$, prohibits direct coupling of the SM fermions with the scalar fields of the triplet. The couplings of the new scalar fields ($H,H^\pm$) with SM fermions are generated after the EWSB. The strength of $H\bar{f}f$ ($f$ are the $up$-,$down$-quarks and charged leptons)  are proportional to $\sin\gamma$. The couplings $H^+ \bar{\nu_l}l^-$ and $H^+ \bar{u}d$ are proportional to $\sin\beta$.
 
\section{Constraints on the hyperchargeless Higgs
triplet model}
\label{sec:constraints}
The parameter space of this model is constrained by theoretical considerations like the absolute vacuum stability, perturbativity, and unitarity of the scattering matrix. In the following, we will discuss these theoretical bounds and the constraints of the Higgs to diphoton signal strength
from the LHC and the electroweak precision measurements.
\subsection{Vacuum stability bounds}
A necessary condition for the stability of the vacuum comes from requiring that the scalar potential is bounded from below, i.e, it should not approach negative infinity along any direction of the field space for large field values. For $h^0,\eta^{0,\pm} \gg v_{1,2}$, the quadratic terms $\mu_{1}^2 |\Phi|^2$, $\frac{\mu_{2}^2}{2} |T|^2$ and $\lambda_4  \Phi^\dagger \sigma^a \Phi T_a$ of the scalar potential in eqn.~\ref{ScalarpotTrip} are negligibly small compared to the other quartic terms, so the scalar potential is given by
\bea
V(h^0, \eta^0,\eta^\pm) &=&  \frac{1}{4} \left[\lambda_1 (h^0)^4 + \lambda_2 (\eta^2+2 \eta^+ \eta^-)^2  + \lambda_3 (h^0)^2 (\eta^2+2 \eta^+ \eta^-)\right].
\eea
The potential can be written in a symmetric matrix with basis $\{ (h^0)^2, (\eta^0)^2, \eta^-\eta^+\}$.
Using the copositivity criteria~\cite{Kannike:2012pe}, one can calculate the required conditions for the absolute stability/bounded from below of the scalar potential. The tree-level scalar potential $V(\Phi,T)\equiv V(h^0, \eta^0,\eta^\pm)$ is absolutely stable if
\beq
\lambda_{1}(\Lambda) \geq 0, \quad  \lambda_{2}(\Lambda) \geq 0, \quad  \lambda_{3}(\Lambda) \geq - 2~\sqrt{ \lambda_{1}(\Lambda)\lambda_{2}(\Lambda)}. \label{stabilitybound}
\eeq
The coupling constants are evaluated at a scale $\Lambda$ using RGEs.
In this study, we use the SM RGEs up to three-loop which have been given in Refs.~\cite{Chetyrkin:2012rz,Zoller:2012cv,Chetyrkin:2013wya,Zoller:2013mra}. The triplet contributions are taken up to two-loop which are presented in Appendix~\ref{App:BetaFunctionsTrip}.
If the quantum corrections are included to the scalar potential, then 
there is a possibility to form a minimum along the Higgs field direction near the Planck mass $\mpl$. For negative $\lambda_1(\Lambda)$ the minimum at the energy scale $\Lambda$ becomes deeper than the EW minimum and vice-versa. In these situations, the above conditions in eqn.~\ref{stabilitybound} become more complicated.
These modifications will be shown in Subsection~\ref{ssec:metastability}.
As $\lambda_3$ gives a positive contribution to the running of $\lambda_2$, $\lambda_2$ remains positive up to the Planck mass $\mpl$. Hence, it is clear that no extra minimum will develop along the new scalar field directions. The sign and the value of $\lambda_3$ can change the Higgs diphoton signal strength and the stability of the EW vacuum. The importance of the sign of $\lambda_3$ will be discussed in subsection~\ref{diphoton} and ~\ref{PhaseDiag}.

\subsection{Perturbativity bounds}
\label{ssec:perturbativity}
To ensure that the radiatively improved scalar potential $V(\Phi, T)$ remains perturbative at any given energy scale ($\Lambda$), one must impose the following conditions,
\beq
\mid \lambda_{1,2,3}\mid \lesssim 4 \pi ~~~{\rm and}~~~\Big |\frac{\lambda_4}{\Lambda}\Big | \lesssim 4 \pi.
\eeq

\subsection{Unitarity bounds}
\label{ssec:unitarity}
Unitarity bound on the extended scalar sectors can be calculated from the scattering-matrix (S-matrix) of different processes. The technique was developed in Refs.~\cite{Lee:1977yc,Lee:1977eg} for the SM and it can also be applied to the HTM ($Y=0$). The S-matrix for the HTM ($Y=0$) consists of different scalar-scalar, gauge boson-gauge boson, gauge boson-scalar scattering amplitudes. Using the Born approximation, the scattering cross-section for any process can be written as
\beq
\sigma = \frac{16 \pi}{s} \sum_{l=1}^\infty ~ (2 l + 1) |a_l(s)|^2,
\eeq
where, $s=4 E_{CM}^2$ is the Mandelstam variable, $E_{CM}$ is the center of mass energy of the incoming particles. $a_l$ is the partial wave coefficients corresponding to specific angular momentum $l$. This leads to the following unitarity constraint: $Re(a_l) < \frac{1}{2}$. At high energy the dominant contribution to the amplitude $a_l$ of the two-body scattering processes $a,b \ra c,d$ comes from the diagram involving the quartic couplings. Far away from the resonance, the other contributions to the amplitude from the scalar mediated $s$-,$t$-, and $u$-channel processes are negligibly small.
Also in the high energy limit, the amplitude of scattering processes involving longitudinal gauge bosons can be approximated by the scalar amplitude in which gauge bosons are replaced by their corresponding Goldstone bosons. For example, the amplitude of $W^+_L W^-_L \ra W^+_L W^-_L$ scattering is equivalent to $G^+ G^- \ra G^+ G^-$.
This is known as equivalence theorem~\cite{Lee:1977eg,equiv}. So to test the unitarity of HTM ($Y=0$), we construct the S-matrix which consists of only the scalar quartic couplings.

The scalar quartic couplings in the physical bases $G^\pm,~G^0~,H^\pm,~h$ and $H$ are complicated functions of $\lambda$'s,$~\gamma,~ {\beta}$.
The $hhhh$ vertex is 6($ \lambda_1 \cos^4\gamma + \lambda_3 \cos^2\gamma \sin^2\gamma + \lambda_2 \sin^4\gamma$).
It is difficult to calculate the unitary bounds in the physical bases.
One can consider the non-physical scalar fields bases, i.e., $G_1^\pm,~\eta^\pm,~ G^0,~h^0$ and $\eta^0$ before the EWSB.
Here the crucial point is that the S-matrix which is expressed in terms of the physical fields can be transformed into a S-matrix for the non-physical fields by making an unitary transformation~\cite{Kanemura:1993hm,Arhrib:2000is}.

Different quartic couplings in non-physical bases are obtained by expanding the scalar potential of eqn.~\ref{ScalarpotTrip} which
are given by,

\begin{minipage}{8cm}
\bea
\{G^0~G^0~G^0~G^0\}&=&6 \lambda_{1},\nn\\
\left\{G_1^+~G_1^+~G_1^-~G_1^-\right\}&=&4 \lambda_{1},\nn\\
\left\{G_1^+~G_1^-~h^0~h^0\right\}&=&2\lambda_{1},\nn\\
\{G^0~G^0~\eta^0~\eta^0\}&=&\lambda_{3},\nn\\
\{h^0~h^0~\eta^0~\eta^0\}&=&\lambda_{3},\nn\\
\left\{G^0~G^0~\eta^{+}~\eta^{-}\right\}&=&\lambda_{3},\nn\\
\left\{h^0~h^0~\eta^{+}~\eta^{-}\right\}&=&\lambda_{3},\nn
\eea
\end{minipage}
\hspace{-3cm}
\begin{minipage}{9cm}
\bea
\left\{G^0~G^0~G_1^+~G_1^-\right\}&=&2 \lambda_{1},\nn\\
\{G^0~G^0~h^0~h^0\}&=&2 \lambda_{1},\nn\\
\{h^0~h^0~h^0~h^0\}&=&6 \lambda_{1}, \nn\\
\left\{G_1^+~G_1^-~\eta^0~\eta^0\right\}&=&
\lambda_{3},~\label{UntiHTM}\\
\{\eta^0~\eta^0~\eta^0~\eta^0\}&=& 6\lambda_{2},\nn\\
\left\{G_1^+~G_1^-~\eta^{+}~\eta^{-}\right\}&=&
\lambda_{3},\nn\\
\left\{\eta^0~\eta^0~\eta^{+}~\eta^{-}\right\}&=&2 \lambda_{2},\nn
\eea
\end{minipage}
\bea
\left\{\eta^{+}~\eta^{+}~\eta^{-}~\eta^{-}\right\}&=&
4\lambda_{2}.\nn ~~~~~~~~~~~~~~~~~~~~
\eea
The full set of these non-physical scalar scattering processes can be expressed as a $16\times16$ S-matrix. This matrix is composed of three submatrices of dimensions $6\times6$, $5\times5$, and $5\times5$ which have different initial and final states.

The first $6\times6$ sub-matrix ${\cal M}_1$ corresponds to scattering processes whose initial and final states are one of these: $h^0~G_1^+,~G^0~G_1^+$ $,~\eta^0~G_1^+,~h^0~G_1^+,~G^0~\eta^+,$ and $\eta^0~\eta^+$. Using the Feynman rules in eqns.~\ref{UntiHTM}, one can obtain ${\cal M}_1$=diag$(~2\lambda_1,~2\lambda_1,
~2\lambda_1$,$~\lambda_3$,~$\lambda_3,~\lambda_3$).

The sub-matrix ${\cal M}_2$ corresponds to scattering processes with one of the following initial and final states: $h^0~G^0,~G_1^+~\eta^-,~\eta^+~G_1^-,
~\eta^0~G^0,$ and $h^0~\eta^0$. Similarly, one can calculate ${\cal M}_2$=diag$(~2\lambda_1,~\lambda_3,
~\lambda_3,~\lambda_3,
~\lambda_3)$.

The third sub-matrix ${\cal M}_3$ corresponds to scattering fields
$(G_1^+~G_1^-,~\eta^+~\eta^-,$ $~\frac{G^0~G^0}{\sqrt{2}},~\frac{h^0~h^0}{\sqrt{2}},$ and $\frac{\eta^0~\eta^0}{\sqrt{2}})$. The factor $\frac{1}{\sqrt{2}}$ is appeared due to statistics of identical particles. ${\cal M}_3$ is given by,
\beq
{\cal M}_3=
\begin{pmatrix}
 4 \lambda_1 & \lambda_3 &  \sqrt{2} \lambda_1 & \sqrt{2}\lambda_1 & \frac{\lambda_3}{ \sqrt{2}} \\
 \lambda_3 & 4 \lambda_2 & \frac{\lambda_3}{ \sqrt{2}} & \frac{\lambda_3}{ \sqrt{2}} &  \sqrt{2} \lambda_2 \\
 \sqrt{2} {\lambda_1} & \frac{\lambda_3}{ \sqrt{2}} & 3 \lambda_1 & \lambda_1 & \frac{\lambda_3}{2} \\
 \sqrt{2} \lambda_1 & \frac{\lambda_3}{ \sqrt{2}} & \lambda_1 & 3 \lambda_1 & \frac{\lambda_3}{2} \\
 \frac{\lambda_3}{ \sqrt{2}} & \sqrt{2} \lambda_2 & \frac{\lambda_3}{2} & \frac{\lambda_3}{2} & 3 \lambda_2
\end{pmatrix}.
\eeq
Eigenvalues of ${\cal M}_3$ are $ 2\lambda_1,~2 \lambda_1,~2 \lambda_2,$ and $\frac{1}{2} \left(6 \lambda_1+5 \lambda_2\pm\sqrt{(6 \lambda_1-5 \lambda_2)^2+12 \lambda_3^2}\right)$.

Unitary constraints of the scattering processes demand that the eigenvalues of the S-matrix should be less than $8\pi$.

\subsection{Bounds from electroweak precision experiments}
Electroweak precision data has imposed severe bounds on new physics models via Peskin-Takeuchi $S, ~T, ~U$ parameters~\cite{Peskin:1991sw}. The additional contributions from this model are given by~\cite{Forshaw:2003kh, Forshaw:2001xq}
{\allowdisplaybreaks
\bea
S  & \simeq & 0, \\
T   &=&  \frac{1}{8\pi} \, \frac{1}{\sin^2\theta_W \cos^2\theta_W} \left[ \frac{M^2_H + M^{2}_{H^\pm}}{M^{2}_{Z}} \;
- \; \frac{2  M^{2}_{H^\pm} M^2_H}{M^{2}_{Z}(M^2_H -  M^{2}_{H^\pm})} \log\left(\frac{M^2_H}{M^{2}_{H^\pm}}\right)
\right] \nonumber \\
 &\simeq & \frac{1}{6\pi} \, \frac{1}{\sin^2\theta_W \cos^2\theta_W} \; \frac{(\Delta M)^{2}}{M^{2}_{Z}}, \\
U   &=&  -\frac{1}{3 \pi} \left( M^4_H\log \left( \frac{M^2_H}{M^{2}_{H^\pm}} \right) \frac{ (3 M^{2}_{H^\pm}-M^2_H)}{(M^2_H-M^{2}_{H^\pm})^3} + \frac{5(M^4_H+M^{4}_{H^\pm})-22 M^{2}_{H^\pm} M^2_H}{6(M^2_H-M^{2}_{H^\pm})^2} \right) \nonumber \\ 
&\simeq &  \frac{\Delta M}{3 \pi M_{H^\pm}},
\eea}
where $\Delta M = M_{H^\pm} - M_{H}$.
$S$ is proportional to $\sin{\beta}$. The experimental value of $\rho$ parameter demands that the triplet VEV $v_2$ to be less than 4 GeV~\cite{Agashe:2014kda}.
Hence, the contributions to the $S$ parameter from the triplet scalar fields are negligible. $M_{H^\pm}$ and $ M_H$ are almost degenerate for $M_{H^\pm,H}\gg M_h$. The contributions to the $T~{\rm and}~U$ parameters from this model are also negligibly small~\cite{Baak:2014ora}.

\subsection{Bounds from LHC diphoton signal strength}
\label{diphoton}
As the dominant production cross-section of $h$ at LHC is coming through gluon fusion, the Higgs to diphoton signal strength $\mu_{\gamma\gamma}$ can be written as
\beq
\mu_{\gamma\gamma} = \frac{\sigma(gg\ra h\ra\gamma\gamma)_{HTM}}{\sigma(gg\ra h\ra\gamma\gamma)_{\rm SM}}= \frac{\sigma(gg\ra h)_{HTM}}{\sigma(gg\ra h)_{SM}}  \frac{Br(h \rightarrow {\gamma\gamma})_{\rm HTM}}{Br(h \rightarrow {\gamma\gamma})_{\rm SM}}.
\eeq
We use the narrow width approximation as $\Gamma_h^{total}/{M_h} \ra 0$. The Higgs $h$ to $f\bar{f}$ and $VV$ ($V$ stands for vector bosons) couplings are proportional to $\cos\gamma$, so $\mu_{\gamma\gamma}$ can be simplified as
\beq
\mu_{\gamma\gamma} = \cos^2\gamma ~\frac{\Gamma^{total}_{h,\rm SM}}{\Gamma^{total}_{h,\rm HTM}}~\frac{\Gamma(h\rightarrow \gamma\gamma)_{\rm HTM}}{\Gamma(h\rightarrow \gamma\gamma)_{\rm SM}}.\label{Hdecay}
\eeq
The charged Higgs $H^\pm$ will alter the decay width of $h\rightarrow\gamma\gamma $, $ Z\gamma$ through one-loop which implies $\Gamma(h\rightarrow \gamma\gamma, Z\gamma) \ll \Gamma^{total}_{h}$. 
Also, if the mass of the extra scalar particles ($HT= H,H^\pm$) happen to be lighter than $M_h/2$, then they
might contribute to the invisible decay of the Higgs boson.
Using the global fit analysis~\cite{Global} that such an invisible branching ratio is less than $\sim 20 \%$. In eqn.~\ref{Hdecay}, the first ratio provides a suppression of $\sim 0.8-1$.
For $M_{H,H^\pm}> M_h/2$, the ratio becomes $\frac{\Gamma^{total}_{h,\rm SM}}{\Gamma^{total}_{h,\rm HTM}}\approx \frac{1}{cos^2\gamma}$. Hence, the Higgs to diphoton signal strength can be written as
\beq 
 \mu_{\gamma\gamma} \approx \frac{\Gamma(h\rightarrow \gamma\gamma)_{\rm HTM}}{\Gamma(h\rightarrow \gamma\gamma)_{\rm SM}}\, .
 \label{mugagahighIT}
\eeq
In HTM, the additional contributions to $\Gamma(h\rightarrow \gamma\gamma)$ at one-loop due to the $H^\pm$ is given by~\cite{Djouadi:2005gj}
\beq
\Gamma(h\rightarrow \gamma\gamma)_{\rm HTM}=\frac{\alpha^2 M_h^3}{ 256\pi^3
v^2}\left|\sum_{f}N^c_fQ_f^2y_f
F_{1/2}(\tau_f)+ y_W F_1(\tau_W)
+Q_{H^{\pm}}^2\frac{v\mu_{{hH^+H^-}}}{
2M_{H^{\pm}}^2}F_0(\tau_{H^{\pm}})\right|^2\
\label{hgaga},
\eeq
where $\tau_i=M_h^2/4M_i^2$. $Q_{f}$, $Q_{H^{\pm}}$ denote electric charges of corresponding particles. $N_f^c$ is the color factor. $y_f$ and
$y_W$ denote the Higgs couplings to $f\bar{f}$ and $W^+W^-$. $\mu_{hH^+H^-}=\{2 {\lambda_4} {\sin\beta} {\cos\beta} {\cos\gamma}  +{\cos\beta}^2 ({\lambda_3} {v_1} {\cos\gamma} +4 {\lambda_2} {v_2} {\sin\gamma} )+{\sin\beta}^2 ({\lambda_4} {\sin\gamma}+{\lambda_1} {v_1} {\cos\gamma} +{\lambda_3} {v_2} {\sin\gamma} )\} \approx \lambda_3 v_{SM}$ stands for the coupling constant of $hH^+H^-$ vertex. The loop functions $F_{(0,\,1/2,\,1)}$  can be found in Ref~\cite{Djouadi:2005gj}.

Recently, the ATLAS~\cite{Aad:2014eha} and CMS~\cite{Khachatryan:2014ira} collaborations have measured the ratio of the diphoton rate $\mu_{\gamma\gamma}$ of the observed Higgs to the SM prediction. The present combined value of $\mu_{\gamma\gamma}$ is $1.14^{+0.19}_{-0.18}$ from these experiments~\cite{Khachatryan:2016vau}.

In $\Gamma(h\rightarrow \gamma\gamma)_{\rm HTM}$ (see eqn.~\ref{hgaga}), a positive $\lambda_3$ leads to a destructive interference between $HT$ and SM contributions and {\it vice versa}.
One can see from the eqn.~\ref{hgaga}, the contribution to the Higgs diphoton channel is proportional to $\frac{\lambda_3}{M_{H^\pm}^2}$.
If the charged scalar mass is greater than $300$ GeV, then the contributions of $H^\pm$ to the diphoton signal is negligibly small.
 \begin{figure}[h!]
 \begin{center}
 \subfigure[]{
 \includegraphics[width=3.in,height=2.8in, angle=0]{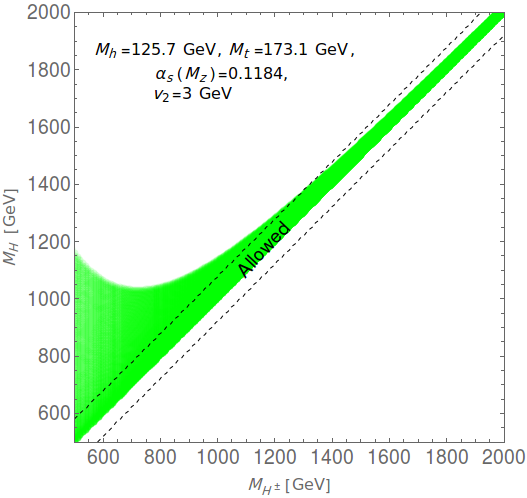}}
 \subfigure[]{
 \includegraphics[width=3.in,height=2.8in, angle=0]{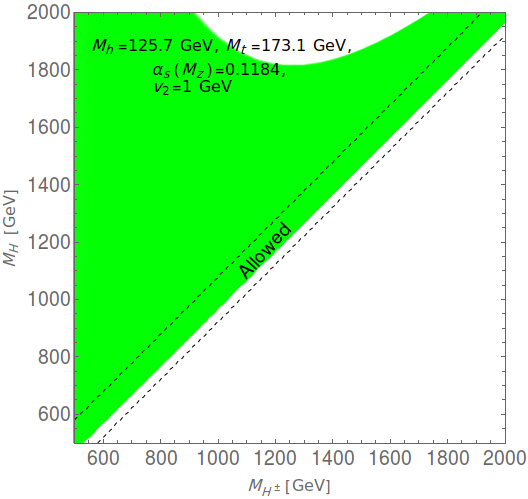}}
 \caption{\label{fig:MTvsMTpm} \textit{The allowed region (green) from the unitarity, perturbativity and absolute stability which is valid up to the Planck mass $\mpl$. The region between the black-dashed line is allowed from the EWPT data at 2$\sigma$.} }
 \end{center}
 \end{figure}

Now we present our results for the central values of the SM parameters such as the Higgs mass $M_h=125.7$ GeV, top mass $M_t=173.1$ GeV, $Z$ boson mass $M_Z=91.1876$ GeV, strong coupling constant $\alpha_s=0.1184$. We take the triplet vev $v_2$, $\lambda_4$ and the other quartic couplings $\lambda_{1,2,3}$ as input parameters. Hence, depending on these parameters the mixing angle $\gamma$ can vary in between $0$ and $\pi/2$. The triplet scalar masses also become arbitrarily heavy. Here, we assume that no new physics shows up below the Planck mass $\mpl$.
We examine the renormalization group (RG) flow of all couplings and establish bounds on the heavy scalar masses under the assumption that the parameters are valid up to the Planck mass $\mpl$.
In this calculation, we use the SM RGEs up to three-loop~\cite{Chetyrkin:2012rz,Zoller:2012cv,Chetyrkin:2013wya,Zoller:2013mra} and the triplet contributions up to two-loop. We first calculate all couplings at $M_t$.
To find their values at $M_t$, one needs to take into account
different threshold corrections up to $M_t$
~\cite{Degrassi:2012ry,Buttazzo:2013uya,
Khan:2014kba,Khan:2015ipa,Sirlin:1985ux,Bezrukov:2012sa}.
Using the RGEs, we evolve all the coupling constants from $M_t$ to the Planck mass $\mpl$. By this procedure we obtain new parameter regions which are valid up to the Planck mass $\mpl$.

We show the allowed region (green) in $M_{H^\pm}-M_H$ plane for this model in Fig. \ref{fig:MTvsMTpm}. We demand that the EW vacuum of the scalar potential remain absolutely stable and do not violate the perturbative-unitarity up to the Planck mass $\mpl$. One can also obtain the parameter spaces, corresponding to the metastable EW vacuum which are visibly small in this plane.
Furthermore, we impose the EWPT constraints on the parameters so that the region between the black-dashed lines survives.

In Fig. \ref{fig:MTvsMTpm}, we show the allowed region for fixed central values of all the SM parameters. In the left panel, we present the plot for the choice of the quartic couplings $\lambda_{2,3}=0.1$ and triplet VEV $v_2=3$ GeV. Whereas in the right panel, we use the value of triplet VEV $v_2=1$ GeV. We vary the quartic coupling $\lambda_1$ and dimensionful mass parameter $\lambda_4$ to calculate the neutral $CP$-even Higgs mass $M_H$, the charged Higgs mass $M_{H^\pm}$ and the mixing angle $\gamma$. These scalar masses increase, whereas mixing angle decreases with $\lambda_4$. We find that the EW vacuum becomes unbounded from below for $\lambda_1\lesssim 0.128$. The theory also violates unitarity bounds for $\lambda_1 \gtrsim 0.238$ before the Planck mass $\mpl$. One can see from the Fig.~\ref{fig:MTvsMTpm} (a), the allowed region becomes smaller for the larger values of heavy scalar masses. In most of the parameter space the running couplings either violate unitary or perturbativity bounds before the Planck mass $\mpl$.

As $\lambda_{2,3}$ stabilize the scalar potential, we will get a wider green region for smaller scalar masses but it will violate the unitarity
bound in the higher mass region. We find that the EW vacuum becomes unbounded from below for the values of the quartic couplings $\lambda_1\lesssim 0.027$ and $\lambda_{2,3} = 0.285$. We also check that the choice of the quartic couplings $\lambda_1 \gtrsim 0.05$ and $\lambda_{2,3} = 0.285$ will violate unitary and perturbativity bounds before the Planck mass $\mpl$.
One can also understand from the expressions of eqns.~\ref{AppMass} that if we decrease the value of $v_2$, the area of allowed region from the stability, unitary and perturbativity bounds will increase. We show the plot in Fig. \ref{fig:MTvsMTpm} (b) for the choice of $v_2=1$ GeV.

If the vacuum expectation value of the scalar triplet becomes zero, then the minimization condition of the scalar potential given in eqn.~\ref{min2} is no longer valid. The mass parameter $\mu_2$ becomes free and the parameter $\lambda_4$ does not play any role in the stability analysis. In the next section, we will show the detailed stability analysis in the presence of extra $\Z_2$-symmetry in this model.
\section{Dark Matter in HTM ($Y=0$)}
\label{sec:ITM}
\begin{figure}[h!]
 \begin{center}
 \subfigure[]{
 \includegraphics[width=3.in,height=2.71in, angle=0]{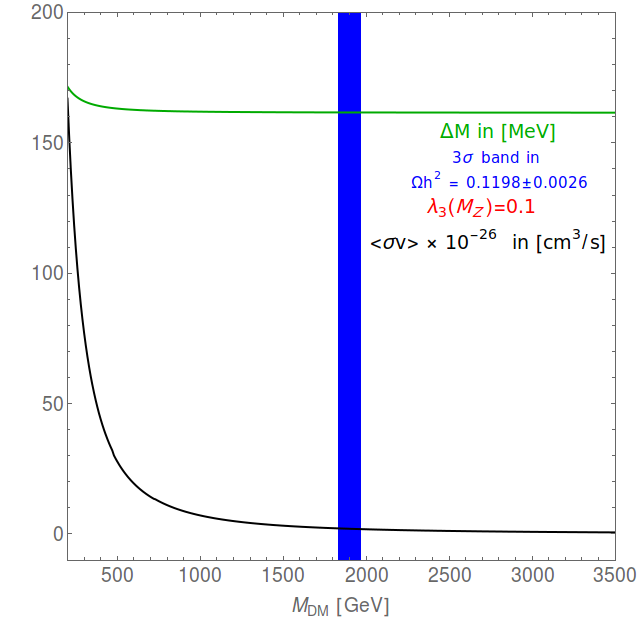}}
 \subfigure[]{
 \includegraphics[width=3.in,height=2.75in, angle=0]{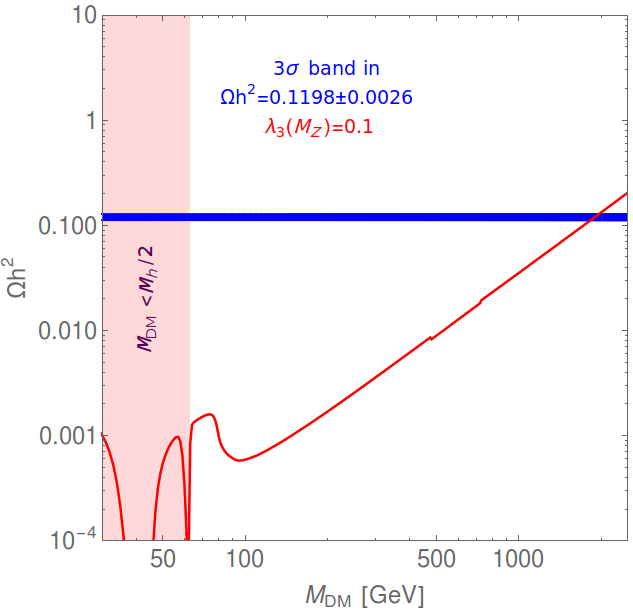}}
 \caption{
The thin blue band corresponds the relic density, $\Omega h^2=0.1198 \pm 0.0026$ $(3\sigma)$ from the combined data of WMAP and Planck~\text{\cite{Ade:2013zuv}}. ($a$) The mass difference $\Delta M$ (green line) and the effective annihilation cross-section (black line) as function of dark matter mass for the portal coupling $\lambda_3(M_Z)=0.10$. ($b$) The relic density $\Omega h^2$ as a function of the DM mass $M_{DM}(\equiv M_H)$ (red line) for $\lambda_3(M_Z)=0.10$.}
\label{fig:relictriplet}
 \end{center}
 \end{figure}

We impose a $Z_2$ symmetry on this model such that the scalar triplet are $odd$ under this transformation, i.e., $T\ra - T$. Whereas SM fields are even under this transformation. In the literature, the HTM including the $\Z_2$-symmetry is known as inert triplet model (ITM)~\cite{Araki:2011hm}. In this model, the term $\lambda_4  H^\dagger \sigma^a \Phi T_a $ is absent in the scalar potential in eqn.~\ref{ScalarpotTrip}, which implies $\lambda_4=0$. The $\Z_2$-symmetry prevents the triplet scalar to acquire a VEV, i.e., $v_2=0$.
The potential can have a minimum along the Higgs field direction only. The EWSB driven by the SM Higgs doublet.
The scalar fields of the triplet do not mix with the scalar fields of SM doublet. 
After the EWSB, the scalar potential in eqn.~\ref{ScalarpotTrip} is then given by
\bea
V(h, H,H^\pm) &=&  \frac{1}{4} \left[ 2 \mu_1^2 (h+v)^2 + \lambda_1 (h+v)^4 +2 \mu_2^2 (H^2+2 H^+ H^-) \right. \nn \\
&& \left. + \lambda_2 (H^2 + 2 H^+ H^-)^2  + \lambda_3 (h+v)^2 (H^2+2 H^+ H^-) \right].
\eea
Here, $v \equiv v_{SM}$ and the mass (see eqn.~\ref{htmy0H0}) of these scalar fields\footnote{ For $v_2=0$, the notation in eqn.~\ref{Tripfield1} $ H \equiv {\eta^{0}} $ and $ H^\pm \equiv {\eta^{\pm}} $ are the physical scalar fields.} $h$, $H$ and $H^\pm$ are given by
\bea
M_{h}^2 &=&  2 \lambda_1 v^2,\nn \\
M_{H}^2 &=& \mu_2^2 +  \frac{\lambda_3}{2} v^2, \\ 
M_{H^\pm}^2 &=& \mu_2^2 + \frac{\lambda_3}{2}  v^2  \,\nn .\label{MassTrip}
\eea
At the tree-level the mass of the neutral scalar $H$ and the charged particles $H^\pm$ are degenerate. If we include one-loop radiative correction, the charged particles become slightly heavier~\cite{Cirelli:2009uv,Cirelli:2005uq} than the neutral ones. The mass difference between them is given by

\beq
\Delta M=(M_{H^\pm}-M_{H})_{1\text{-}loop}=\frac{\alpha M_{H}}{4\pi}\Big[f\Big(\frac{M_W}{M_{H}}\Big) -c_W^2 f\Big(\frac{M_Z}{M_{H}}\Big)\Big]
\label{massdifftrip},
\eeq
with,
$f(x)=-\frac{x}{4}\Big\{ 2 x^3 ~ {\rm log}(x)+(x^2-4)^{\frac{3}{2}}~ {\rm log}\left( \frac{x^2-2-x\sqrt{x^2-4}}{2} \right)\Big\}$.
It has been shown in Refs~\cite{Cirelli:2009uv,Cirelli:2005uq} that the mass splitting between charged and neutral scalars remains $\sim150$ MeV for $M_H = 0.1-5$ TeV. In Fig.~\ref{fig:relictriplet} (a), we show the variation $\Delta M$ (green line) with the $M_H$ ($\equiv M_{DM}$) mass.
As the $\Z_2$-symmetry also prohibits the couplings of an $odd$ number of scalar fields of the triplet with the SM particles, $H$ can serve as a viable DM candidate which may saturate the measured DM relic density of the Universe.
In this work, we use the software package {\tt FeynRules}~\cite{Alloul:2013bka} along with {\tt micrOMEGAs}~\cite{Belanger:2010gh, Belanger:2013oya} to calculate the relic density of the DM. As $\Delta M$ is very small, the 
effective annihilation cross-section is dominated by the co-annihilation channels $H H^\pm \ra {\rm SM ~particles}$~\cite{Griest:1990kh}.
Although it is dominated by the co-annihilation channel, we need a very small Higgs portal coupling $\lambda_3$ to obtain the correct relic density.
The effective annihilation cross-section (see the black line in Fig.~\ref{fig:relictriplet} (a)) decreases rapidly with  $\Delta M$ for the DM mass below $500$ GeV and becomes $\sim 10^{-26}$ $\text{cm}^3 s^{-1}$ around $M_{DM}=2000$ GeV. We obtain the relic density in the right ballpark.

In Fig.~\ref{fig:relictriplet} (b), we present the plot for the relic density as a function of DM mass for the fixed Higgs portal coupling $\lambda_3(M_Z)=0.10$. The light red band is excluded from the Higgs invisible decay width~\cite{Belanger:2013xza}. There are two deep region in the relic density band (red line). First one is situated near the DM mass $M_{DM}\approx 45$ GeV. It is due the resonance of the $s$-channel $H H^\pm \rightarrow \text{SM~fermions}$ processes which is mediated by the vector bosons $W^\pm$. The second one is situated near the DM mass $M_{DM}\approx M_h/2$ for the Higgs-mediated $H H \rightarrow \text{SM~fermions}$ processes. There is another shallower region located around the DM mass $M_{DM}=100$ GeV, which is due to the dominant contributions coming from $H H^\pm, H H \rightarrow \text{gauge~bosons}$ channels.

For 500 GeV, we find that the total cross-section $\left\langle \sigma v \right\rangle \sim 10^{-25}$ ${\rm cm^3 s^{-1}}$, so the relic density becomes $\sim 0.01$. In this region, the dominant channel are $H,H^\pm\rightarrow Z W^\pm, \gamma W^\pm$ ($\sim 35\%$, $\sim 10\%$) and $H^\pm,H^\pm\rightarrow Z W^\pm$ ($\sim 25\%$). We also check that the smaller dark matter mass along with the Higgs portal coupling $\lambda_3$ (within the perturbative limit) does alter the relic density only in the $third$ decimal place. If we increase the DM masses, then the effective annihilation cross-section decrease. It is mainly due to the mass suppression. We get a DM relic density in the right ballpark for DM masses greater than $1.8$ TeV. 

One can see that the mass splitting $\Delta M$ attains saturation for $M_{DM} > 700$ GeV. Hence, the relic density is mainly regulated by the Higgs-mediated $s$-channel processes, although the contributions are small. We check that the Higgs portal coupling $\lambda_3$ can be varied in between $0$ to $1$ for the DM mass $1850$ GeV to $2200$ GeV to get the right relic density. For example, we obtain the relic density $\Omega h^2 =0.1198$ for $\lambda_3=0.001$ and $M_{DM}=1894.5$ GeV. We get the same relic density for $\lambda_3=0.8$ and $M_{DM}=2040$ GeV. However, the running couplings will violate the unitary and perturbativity bounds for $\lambda_3 \gtrsim 0.6$.

Non-observation of DM signals in direct detection experiments at XENON\,100~\cite{Aprile:2011hi,Aprile:2012nq}, LUX~\cite{Akerib:2013tjd} and LUX-2016~\cite{Akerib:2016vxi} put severe restrictions~\cite{Ayazi:2014tha} on the Higgs portal coupling $\lambda_3$ for a given DM mass. In this model, we check the parameter regions which are satisfying the relic density, are allowed by the recent LUX-2016~\cite{Akerib:2016vxi} and XENON1T-2017~\cite{Aprile:2017iyp} data.


\subsection{Metastability in ITM ($Y=0$)}
As in the SM, the EW vacuum is metastable, it is important to explore if ITM has any solution in its reserve. As the scalar WIMP $H$ protected by $\Z_2$-symmetry can serve as viable DM candidate, it is interesting to explore if they help prolong the lifetime of the Universe. The effective Higgs potential gets modified in the presence of these new extra scalars.

One-loop effective Higgs potential in $\MS$ scheme and the Landau gauge is given by   
\beq
V_1^{{\rm SM}+{\rm IT}}(h)= V_1^{\rm SM}(h) + V_1^{\rm IT}(h),
\eeq
where~\cite{Casas:1994qy, Altarelli:1994rb, Casas:1994us, Casas:1996aq, Quiros:1997vk}
\beq
V_1^{\rm SM}(h)=\sum_{i=1}^5 \frac{n_i}{64 \pi^2} M_i^4(h) \left[ \ln\frac{M_i^2(h)}{\mu^2(t)}-c_i\right] \, .
\eeq

$n_i$ is the number of degrees of freedom and $M_i^2(h)= \kappa_i(t)\, h^2(t)-\kappa_i^{\prime}(t)$.
$n_i$, $c_i$, $\kappa_i$ and $\kappa_i^{\prime}$ can be found in Eqn.~(4) in Ref.~\cite{Casas:1994qy}. $t$ is a dimensionless parameter which is expressed in terms of the running
parameter $\mu(t)=M_Z \,exp(t).$

The contributions to the effective Higgs potential from the new scalars ($H,H^\pm$) of the inert scalar triplet are given by~\cite{Forshaw:2003kh}
\beq
V_1^{\rm IT}(h)= \sum_{j=H,H^+,H^-} \frac{1}{64 \pi^2} M_j^4(h) \left[ \ln\left(\frac{M_j^2(h)}{\mu^2(t)} \right)- \frac{3}{2} \right], 
\eeq
where, 
$
M_j^2(h)=\frac{1}{2} \,\lambda_{j}(t) \, h^2(t)+\mu_2^2(t)
$,
with $\lambda_{H,H^\pm}(t)= \lambda_{3}(t)$.
In the present work, in the Higgs effective potential, SM contributions are taken up to two-loop level~\cite{Degrassi:2012ry, Buttazzo:2013uya,Ford:1992pn,Martin:2001vx} and the IT scalar contributions are considered up to one-loop only~\cite{Forshaw:2003kh}.

For $h\gg v$, the quantum corrections to the Higgs potential are reabsorbed in the effective running coupling $\lambda_{\rm 1,eff}$ such that the effective potential becomes
\beq
V_{\rm eff}^{{\rm SM}+{\rm IT}}(h) \simeq \lambda_{\rm 1,eff}(h) \frac{h^4}{4}\, ,
\eeq
with
\beq
\lambda_{\rm 1,eff}(h) = \lambda_{\rm 1,eff}^{\rm SM}(h) +\lambda_{\rm 1,eff}^{\rm IT}(h)\, ,
\label{efflam}
\eeq
where, the expression of $\lambda_{\rm 1,eff}^{\rm SM}(h)$ up to two-loop quantum corrections can be found in Ref.~\cite{Buttazzo:2013uya} and 
$
 \lambda_{\rm 1,eff}^{\rm IT}(h)= e^{4\Gamma(h)} \left[\frac{3 \lambda_3^2}{256 \pi^2}  \left(\ln\left(\frac{\lambda_3}{2}\right)-\frac{3}{2}\right ) \right]\,
$,
with
$
\Gamma(h)=\int_{M_t}^{h} \gamma(\mu)\,d\ln\mu
$.
The wave function renormalization of the Higgs field is taken into account by the anomalous dimension $\gamma(\mu)$.
Here, all running coupling constants are evaluated at $\mu=h$, ensuring the potential remains within the perturbative domain.

We first calculate all couplings with the threshold corrections~\cite{Degrassi:2012ry,Buttazzo:2013uya,
Khan:2014kba,Khan:2015ipa,Sirlin:1985ux,Bezrukov:2012sa} at $M_t$.
Then we evolve all the couplings up to the Planck mass $\mpl$ using our own computer codes incorporating the RG equations. Here, the SM effects in the RGEs are taken up to three-loop~\cite{Chetyrkin:2012rz,Zoller:2012cv,Chetyrkin:2013wya,Zoller:2013mra} and IT contributions are considered up to two-loop (see appendix~\ref{App:BetaFunctionsTrip}).
\begin{table}[h!]
\begin{center}
    \begin{tabular}{| c | c | c | c | c | c | c | c |}
    \hline
      & $\lambda_1$ & $\lambda_2$ &  $\lambda_3$\\
\hline
   ~$M_{t}$  ~&~ 0.127054 ~&~ 0.10 ~&~ 0.10~\\
            \hline
   $\mpl$ & $-$0.00339962 & 0.267706 & 0.206306\\
              \hline
    \end{tabular}
    \caption{\textit{A set of values of all quartic coupling constants at  $M_t$ and $\mpl$ for $M_{DM}=1897$ }\text{GeV.}}
    \label{table1}
\end{center}
\end{table}

\begin{figure}[h!]
 \begin{center}{
 \includegraphics[width=3.5in,height=3.0in, angle=0]{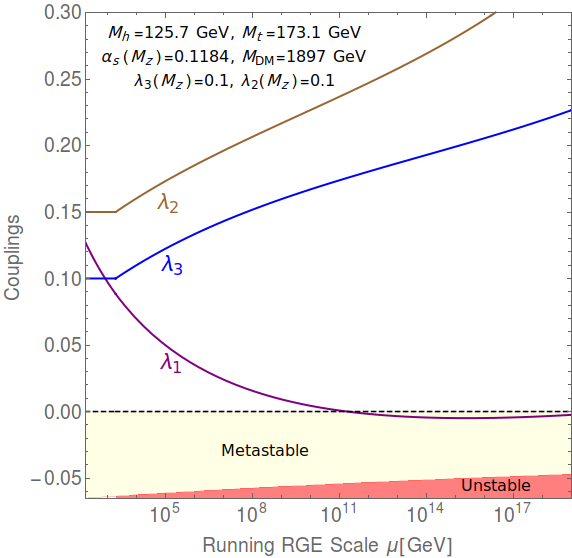}}
 \caption{\label{fig:SMInert} \textit{RG evolution of the couplings $\lambda_{1,2,3}$ for the set of parameters in Table~\ref{table1} with DM mass $M_{DM}=1897$} \text{GeV}. }
 \end{center}
 \end{figure}

We choose a specific benchmark point $M_{DM}(\equiv M_H)=1897$ GeV,
$M_h=125.7$~GeV and $\alpha_s\left(M_Z\right)=0.1184$ such that it can give the right DM density of the Universe. The corresponding values of all quartic couplings $\lambda_{1,2,3}$ at $M_t=173.1$~GeV and $\mpl = 1.2 \times 10^{19}$~GeV are presented in Table~\ref{table1}. For this benchmark point, we show the evolution of the running of the quartic couplings $(\lambda_{1,2,3})$ in Fig.~\ref{fig:SMInert}.
We find that for this specific choice of benchmark point with the top mass\footnote{As the $\beta$function of the Higgs quartic coupling, $\lambda_1$ contains $-\frac{6 y_t^4}{16 \pi^2}$ (see eqn.~\ref{betal_11}), the values of the Higgs quartic couplings $\lambda_{1}$ at very high energies are extremely sensitive to $M_t$.} $M_t=173.1$~GeV and the central values of other SM parameters leads to a metastable EW vacuum. It implies that the $\beta$function of the Higgs quartic coupling $\lambda_1$ becomes zero at very high energy scale and remains positive up to the Planck mass $\mpl$. We find that a deeper minimum is situated at that high energy scale before the Planck mass $\mpl$.
We also check that the EW vacuum remains metastable (one-sided) for the quartic coupling $\lambda_2\leq 0.1$, Higgs portal coupling $\lambda_3 \leq 0.15$ and DM mass $M_{DM} \geq 1900$ GeV. We obtain the stable ($>99.99\%$ confidence level, one-sided) EW vacuum for the choice of the parameters $\lambda_2=0.1$, $\lambda_3=0.3$ and $M_{DM}=1915$ GeV. The running couplings will violate the unitary and perturbativity bounds for $\lambda_3 \gtrsim 0.6$.
In the following subsections, we will discuss the metastability of the EW vacuum of the scalar potential. 
\subsection{Tunneling Probability}
Using the experimentally measured values of the SM parameters at the EW scale, when analyzing the SM scalar potential at higher energy scales, one encounters the so-called metastability of EW vacuum~\cite{Degrassi:2012ry,Buttazzo:2013uya,Khan:2014kba,
Khan:2015ipa,Masina:2012tz}.
Since a second (true) minimum, deeper than the EW minimum, is situated near the Planck mass, there exists a non-zero probability that the EW minimum will tunnel into the second minimum.
The tunneling probability of the EW vacuum to the true vacuum at the present epoch can be expressed as~\cite{Coleman:1977py, Isidori:2001bm, Buttazzo:2013uya}
\beq
{\cal P}_0=0.15 \frac{\Lambda_B^4}{H^4} e^{-S(\Lambda_B)},
\label{prob}\\
\eeq
where, $S(\Lambda_B)$ is the minimum action of the Higgs potential of bounce of size $R=\Lambda_B^{-1}$ and is given by
\beq
S(\Lambda_B)=\frac{8\pi^2}{3|\lambda_1(\Lambda_B)|}\, .
\label{action}\\
\eeq
It becomes minimum when $\lambda_1(\Lambda_B)$ is minimum, \ie $\beta_{\lambda_1}(\Lambda_B)=0$. In this work, we neglect loop~\cite{Isidori:2001bm} and gravitational corrections~\cite{Coleman:1980aw,Isidori:2007vm} to the action as in Ref.~\cite{Khan:2014kba,Khan:2015ipa}. Finite temperature also affects to EW vacuum stability~\cite{Isidori:2001bm,Rose:2015lna,Espinosa:1995se}. In this work, we consider field theory in the zero-temperature limit.

\begin{figure}[h!]
 \begin{center}
\subfigure[]{
 \includegraphics[width=2.75in,height=2.75in, angle=0]{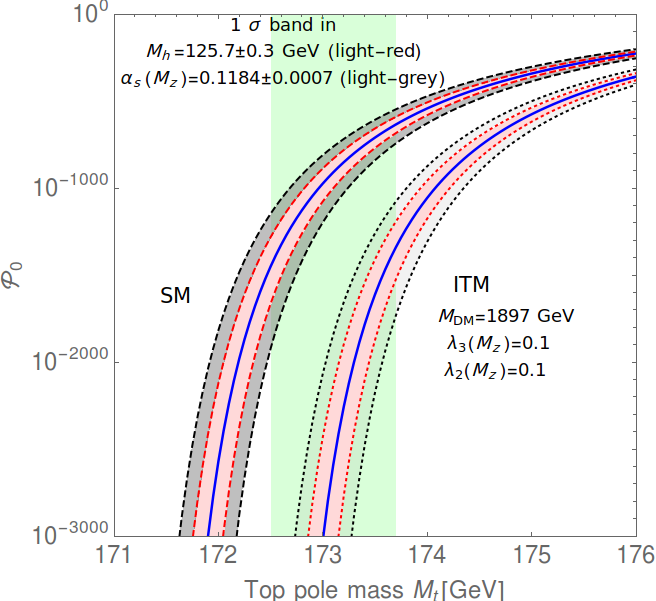}}
 \hskip 15pt
 \subfigure[]{
 \includegraphics[width=2.75in,height=2.75in, angle=0]{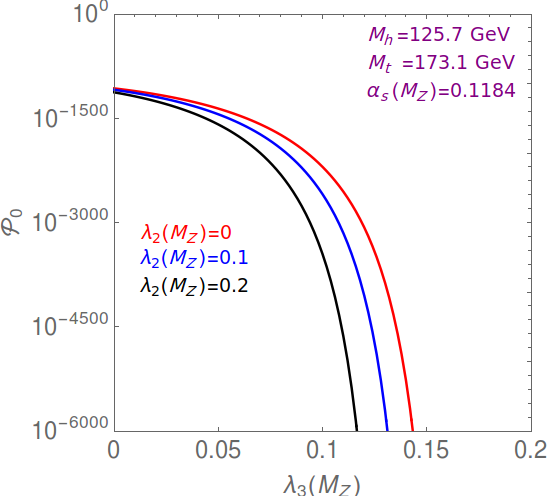}}
 \caption{\label{fig:Tun} \textit{ (a) Tunneling probability ${\cal P}_0$ dependence on  $M_t$. The left band (between dashed lines) corresponds to {\rm SM}. The right one (between dotted lines) is for $IT$ model for DM mass $M_H=1897$~GeV. Dark matter constraints are respected for these specific choice of parameters. Light-green band stands for $M_t$ at $\pm 1\sigma$. (b) ${\cal P}_0$ is plotted against the Higgs DM coupling $\lambda_3(M_Z)$ for different values of $\lambda_2(M_Z)$.}}
 \end{center}
 \end{figure}

In the ITM, the additional scalar fields give a positive contribution to 
$\beta_{\lambda_1}$ (see Eqns.~\ref{betal_11}, \ref{betal_12}). Due to the presence of these extra scalars, a metastable EW vacuum goes towards the stability, i.e., the tunneling probability ${\cal P}_0$ becomes smaller. 
We first calculate the minimum value of $\lambda_{1,\rm eff}$ of eqn.~\ref{efflam}. Putting this minimum value in eqn.~\ref{action}, we compute the tunneling probability ${\cal P}_0$. As the stability of the EW vacuum is very sensitive to the top mass $M_t$, we show the variation of tunneling probability ${\cal P}_0$ as a function of $M_t$ in Fig.~\ref{fig:Tun}$\left(a\right)$.
The right band in Fig.~\ref{fig:Tun}$\left(a\right)$ corresponds to the tunneling probability for our benchmark point.
We present ${\cal P}_0$ for the SM as the left band to see the effect of the additional IT scalar. We also display
1$\sigma$ error bands in $\alpha_s$ (light-grey) and $M_h$ (light-red). One can see from this figure that the effect of $\alpha_s$ on the tunneling probability is more than the effect of $M_h$.  
To see the effect of the ITM parameter spaces,
we plot ${\cal P}_0$ as a function of the Higgs portal coupling $\lambda_3(M_Z)$ in Fig.~\ref{fig:Tun}$\left(b\right)$ for different choices of $\lambda_2(M_Z)$. We keep the fixed central values of all SM parameters. 
Here, DM mass $M_{DM}$ is also varied with $\lambda_3$ to get the DM relic density $\Omega h^2=0.1198$.

The additional IT scalar fields in the IT model improve the stability of the EW vacuum as

\begin{itemize}
\item
If $0>\lambda_1(\Lambda_B)>\lambda_{\rm 1,min}(\Lambda_B)$, then the vacuum is metastable. 
\item
If $\lambda_1(\Lambda_B)<\lambda_{\rm 1,min}(\Lambda_B)$, then the vacuum is unstable. 
\item
If $\lambda_2<0$, the potential is unbounded from below along the $H$ and $H^\pm$-direction. 
\item
If $\lambda_3(\Lambda_{\rm I})<0$, the potential is unbounded from below along a direction in between $H$ and $h$ also $H^\pm$ and $h$.
\end{itemize}
In the above $\lambda_{\rm 1,min}(\Lambda_B)=\frac{-0.06488}{1-0.00986 \ln\left( {v}/{\Lambda_B} \right)}$ and $\Lambda_{\rm I}$ represents any energy scale for which $\lambda_1$ is negative~\cite{Khan:2014kba,
Khan:2015ipa}.
\label{ssec:metastability}

\subsection{Phase diagrams} 
\label{PhaseDiag} 
In order to show the explicit dependence of the electroweak stability for different parameters of the ITM, we present various kinds of phase diagrams.
 \begin{figure}[h!]
 \begin{center}
 \subfigure[]{\includegraphics[width=2.7in,height=2.7in, angle=0]{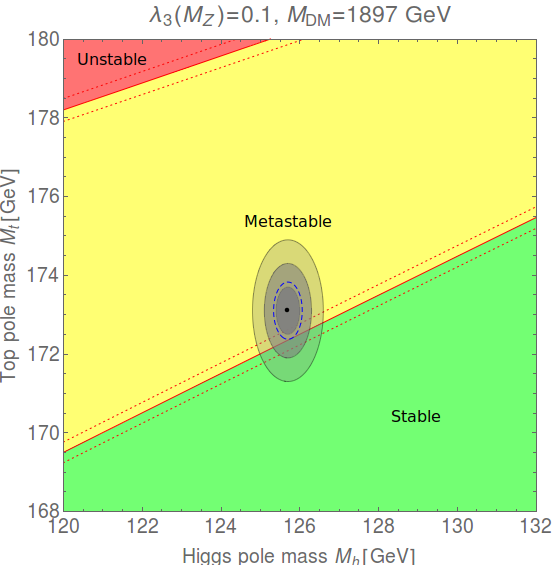}}
 \hskip 15pt
  \subfigure[]{
 \includegraphics[width=2.7in,height=2.7in, angle=0]{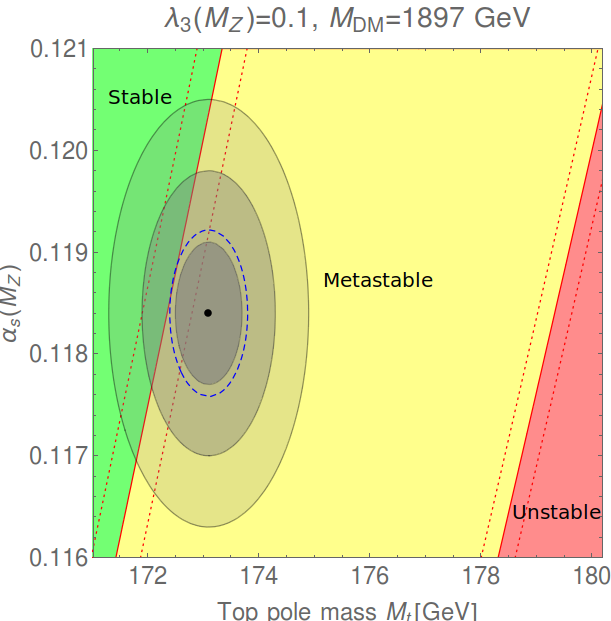}}
 \caption{\label{fig:Mt_MhIT} \textit{Phase diagrams in (a) $M_h - M_t$ plane and (b) $M_t-\alpha_s(M_Z)$ plane {\rm ITM}. Regions of 
absolute stability~(green), metastability~(yellow), instability~(red) of the EW vacuum are also marked. The gray zones represent error ellipses at $1$, $2$ and  $3\sigma$. The three boundary lines (dotted, solid and dotted red) correspond to $\alpha_s(M_Z)=0.1184 \pm 0.0007$.
} }
 \end{center}
 \end{figure} 
In Fig.~\ref{fig:Mt_MhIT} (a), we calculate the confidence level for our bench mark points $M_{DM}=1897$ GeV, $\lambda_2(M_Z)=0.10$ and $\lambda_3(M_Z)=0.10$ by drawing an ellipse passing through the stability line $\lambda=\beta_{\lambda}=0$ in $M_t-M_h$ plane. If the area of the ellipse is $\chi$ times the area of the ellipse which represents the $1~\sigma$-error in the same plane. This factor $\chi$ is the confidence level of the stability of EW vacuum.
We develop a proper method to calculate this factor and the tangency point for the stability line. In this case, the confidence level of metastability is decreased (one-sided) with $\alpha_s(M_Z)$, i.e., the EW vacuum moves towards the stability region.
We obtain the similar factor in the $\alpha_s(M_Z)-M_t$ plane. In this case, the confidence level decreases with $M_h$.
One can see from the phase diagrams in Fig.~\ref{fig:Mt_MhIT} that the stable EW vacuum is excluded at 1.2 $\sigma$ (one-sided).

 \begin{figure}[h!]
 \begin{center}{
 \includegraphics[width=2.7in,height=2.7in, angle=0]{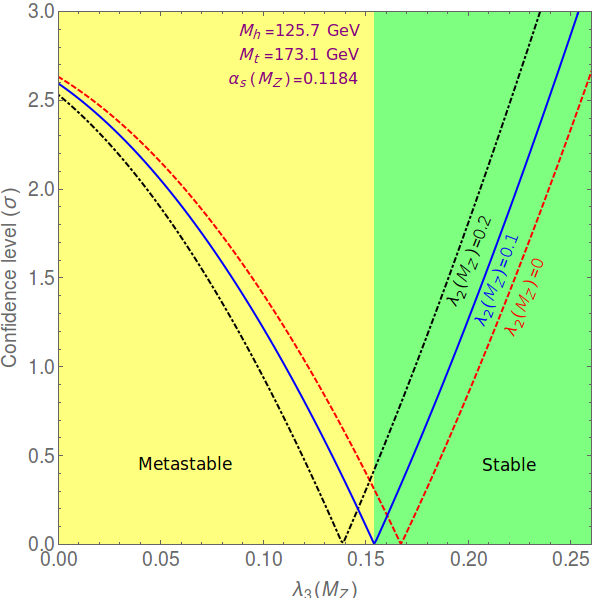}}
 \caption{\label{fig:confidenceIT} \textit{Dependence of confidence level  at which the EW vacuum stability is excluded (one-sided) or allowed on $\lambda_3(M_Z)$ and $\lambda_2(M_Z)$ in {\rm ITM}. Regions of absolute stability (green) and metastability (yellow) of EW vacuum are shown for $\lambda_2(M_Z)=0.1$.     
 } }
 \end{center}
 \end{figure}

If the ITM is valid up to the Planck mass which also saturates the DM abundance of the Universe then the confidence level vs $\lambda_3(M_Z)$ phase diagram becomes important to realize where the present EW vacuum is residing.
In Fig.~\ref{fig:confidenceIT}, we vary the DM mass with $\lambda_{3}(M_Z)$
to keep the relic density at $\Omega h^2 = 0.1198$.
One can see that the EW vacuum approaches the stability with larger values of $\lambda_{2,3}(M_Z)$. The EW vacuum becomes absolutely stable for $\lambda_3(M_Z)\geq 0.154$ and $\lambda_2(M_Z)\approx 0.10$ (see blue line in Fig.~\ref{fig:confidenceIT}).
We show this phase diagram for central values of the SM parameters. 
Moreover, if we increase the top mass and/or decrease the Higgs mass along with $\alpha_s(M_Z)$ then 
the size of the region corresponding to the metastable EW vacuum will be increased.
We see that the conditions, the DM mass $M_{DM}\geq1912$ GeV, $\lambda_{3}(M_Z)\geq0.31$ and $\lambda_{2}(M_Z)\geq0.1$ are required to stabilize the EW vacuum for $M_t=174.9$ GeV, $M_h=124.8$ GeV and $\alpha_s(M_Z)=0.1163$.

\begin{figure}[h!]
 \begin{center}{
 \includegraphics[width=3.0in,height=2.7in, angle=0]{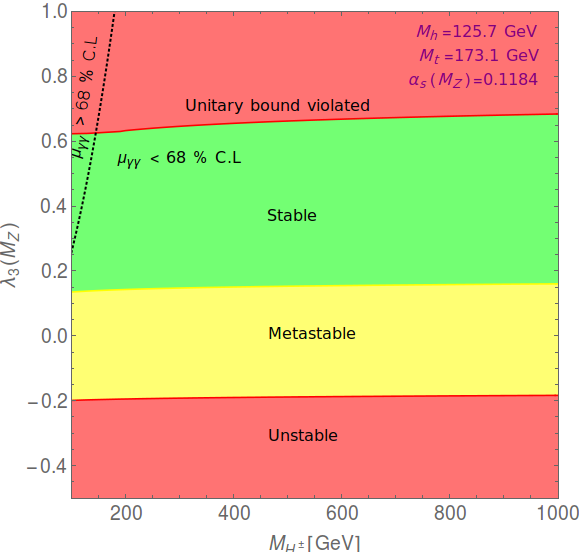}}
 \caption{\label{fig:Rgammagamma} \textit{Phase diagram in $\lambda_3(M_Z)-M_{H^\pm}$ plane in {ITM}. Right side of the black-dotted line is allowed from the signal strength ratio of $\mu_{\gamma\gamma}$ within 68$\%$ confidence level and the left side is excluded at 1$\sigma$. In the metastable region, the Higgs portal coupling $\lambda_3(M_Z)$ is negative, however, beyond the scale $\Lambda_{B}$ it is greater than zero.} }
 \end{center}
 \end{figure}

In Fig.~\ref{fig:Rgammagamma}, we show the allowed parameter spaces in $\lambda_3(M_Z)-M_{H^\pm}$ plane for central values of SM parameters and $\lambda_2(M_Z)=0.1$. The lower (red) region is excluded since the scalar potential becomes unbounded from below along the direction in between $H^\pm$ and $h$.
In this region, the effective Higgs quartic coupling is negative and at the same time $\lambda_3$ remains negative up to the Planck mass $\mpl$. We obtain the parameter space with negative $\lambda_3(M_Z)$ which is also allowed from metastability.
In this case, $\lambda_3$ becomes positive at the scale $\Lambda_{B}$ and remains positive up to the Planck mass $\mpl$. The EW vacuum is absolutely stable in the green region. The upper red region violates unitary bounds. The right-side of the black dotted line are allowed from $\mu_{\gamma\gamma}$ at 1$\sigma$.

\section{Inflation in HTM($Y=0$)}
\label{sec:infl}

Observations of super-horizon ansiotropies in the CMB data, measured by various experiments such as WMAP, Planck have established that the early Universe underwent a period of rapid expansion. This is known as inflation. This can solve a number of cosmological problems such as the horizon problem, the flatness problem and the magnetic monopole problem of the present Universe.
If the electroweak vacuum is metastable then the Higgs is unlikely to play the role of
inflaton \cite{Bezrukov:2007ep,Bezrukov:2008ut,Bezrukov:2008ej,Bezrukov:2009db,Barvinsky:2008ia,Barvinsky:2009fy,DeSimone:2008ei,GarciaBellido:2008ab,Park:2008hz} in the SM. Therefore, extra new degrees of freedom are needed in addition to the SM ones to explain inflation in
the early Universe \cite{Lerner:2009xg,Lebedev:2011aq,Chakravarty:2014yda,Chakravarty:2016fin,Chakravarty:2016avd,Ellis:2017upx}.

Here, we study an extension of the Higgs sector with a real triplet scalar $T$ in the presence
of large couplings $\zeta_{h,H}$ to Ricci scalar curvature $R$. This theory can explain inflation in the early Universe at the large field values in the scale invariance Einstein frame.  

In this model, the action of the fields in Jordon frame is given by
\beq
S_j = \int \sqrt{-g} d^4x\left[ {\cal L}_{SM} +  \frac{1}{2} (\partial_\mu \Phi)^\dagger (\partial_\mu \Phi)+\frac{1}{2}(\partial_\mu T)^\dagger (\partial_\mu T) - \zeta_h R |\Phi|^2 - \zeta_H R |T|^2 -V(\Phi,T)\right],
\label{Jorda}
\eeq
In the present work, we consider $H$ as an inflaton. The Higgs $h$ can also act as an inflaton for the stable EW vacuum. In order to calculate the infaltionary observables such as the tensor-to-scalar ratio $r$, spectral index $n_s$ and running of the spectral index $n_{rs}$, we perform a conformal
transformation from Jordon frame to the Einstein frame so that the non-minimal coupling $\zeta_H$ of scalar field to the Ricci scalar disappears.

The transformations is given by \cite{Kahlhoefer:2015jma}
\beq
\tilde{g}_{\mu\nu} = \Omega^2 g_{\mu\nu},~~~~\Omega=\sqrt{1+\zeta_H \frac{H^2}{\mpl^2}}
\eeq

\begin{figure}[h!]
 \begin{center}
 {
 \includegraphics[width=3.2in,height=2.4in, angle=0]{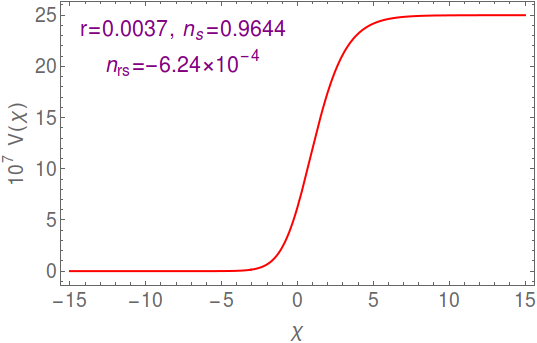}}
 \caption{\label{fig:inflation} \textit{ Inflation potential in the Planck unit for $\zeta_H=1$ and $\lambda_2=10^{-9}$.} }
 \end{center}
 \end{figure}

The action of eqn.~\ref{Jorda} in Einstein frame can be written as
\beq
S = \int \sqrt{-g} d^4x\left[ \frac{1}{2}(\partial_\mu \chi)^\dagger (\partial_\mu \chi) -V(\chi)\right],
\eeq
where,
\beq
\frac{d\chi}{dH} = \sqrt{\frac{\Omega^2 \mpl^2+ 6 \zeta_H^2 H^2}{\Omega^4 \mpl^2}}
\eeq
The scalar potential $V(\chi)$ is then given by
\beq
V(\chi) = \lambda_2 \frac{\mpl^4}{4 \zeta_H^2} \left( 1+ exp\left( -\sqrt{\frac{2 \chi}{3 \mpl}}\right)\right)^{-2}.
\eeq
We plot this potential in Fig.~\ref{fig:inflation} for the choice of bench mark point $\zeta_H=1$ and $\lambda_2=10^{-9}$.
One can also get the same plot for the parameters $\zeta_H=10^4$ and $\lambda_2=0.1$. However, this choice of the parameters violate the unitary bound. One can see that the potential have the ability to explain slow-roll inflation.

One can define the slow-roll parameters $\epsilon,\eta~ \text{and}~ \zeta$ in terms of the potential as
\beq
\epsilon =  \frac{1}{2} \left( \frac{1}{V} \frac{dV}{d\chi} \right)^2,~~~
\eta = \frac{1}{V} \frac{d^2V}{d\chi^2},~~~\text{and}~~~
\zeta = \frac{1}{V^2} \frac{dV}{d\chi} \frac{d^3V}{d\chi^3} \nn.
\eeq
The inflationary observable quantities such as the tensor-to-scalar ratio $r$, the spectral index $n_s$ and the running of the spectral index $n_{rs}$ are defined as
\beq
r=16 \epsilon,~~~n_s = 1 -6 \epsilon + 2 \eta,~~~\text{and}~~~ n_{rs} = -2 \zeta -24 \epsilon ^2+16 \eta  \epsilon
\eeq
and the number of $e$-folds is given by
\beq
N = \int^{\chi_{end}}_{\chi_{start}}~ \frac{V}{dV/d\chi} d\chi
\label{eq:efold}
\eeq
where $\chi_{start}$ ($\chi_{end}$) is the initial (final) value when inflation starts (ends). At $\chi_{start}$, $\epsilon$ is $one$. We calculate the $\chi_{end}$ form the above eqn.~\ref{eq:efold} for $N=60$.

At the end of inflation, we get
\beq
r=0.0037,~~ n_s=0.9644,~~~\text{and}~~ n_{rs}=- 6.24 \times 10^{-4}
\eeq

which is allowed by the present experimental data at 1$\sigma$ \cite{Ade:2015xua,Ade:2015lrj}. Hence, the neutral component of the triplet scalar can simultaneously serve as a inflaton and dark matter particle as well.

\section{Discussion and Conclusions}
\label{sec:conclu}
The measurements of the properties
of the Higgs-like scalar boson detected at the Large Hadron Collider on 4th July 2012
are consistent with the minimal choice of the scalar sector.
But the experimental data of the
Higgs signal strengths and the uncertainties in the measurement of other standard model
parameters still allow an extended scalar sector.
We have taken an extra hyperchargeless scalar triplet as a new physics.
First, we have considered that the extra neutral $CP$-even component of the scalar triplet has also participated in the EWSB. We have shown the detailed structure of the tree-level scalar potential and mixing of the scalar fields.
We have also discussed the bounds on the VEV ($v_2$) of the neutral $CP$-even component of the scalar triplet from the $\rho$-parameter.
To the best of our knowledge
the full expressions of unitary bounds on the quartic couplings of the scalar potential in this model have not yet been presented in the literature.
We have shown these unitary bounds in this model.
As the SM gauge symmetry $SU(2)_L$ prohibits the coupling of SM neutrinos with the neutral $CP$-even component ($\eta^0$) of the scalar triplet, the model does not give neutrino masses. But the model is still interesting as it can play the role in improving the stability of the Higgs potential. 
We have taken into account various threshold corrections to calculate all the couplings at $M_t$. Then
using three-loop SM RGEs and two-loop triplet RGEs, we have evolved all the couplings up to the Planck mass $\mpl$.
We have shown the allowed region in $M_{H^\pm}-M_H$ plane. We have demanded that the EW vacuum of the scalar potential remain absolutely stable and do not violate the perturbative-unitarity up to the Planck mass $\mpl$. We have discussed the constraints on the parameter spaces from the recent LHC $\mu_{\gamma\gamma}$ and $\mu_{Z\gamma}$ data.
Furthermore, only a very small region of the parameter space is shown to survive on imposing the EWPT constraints.

Various kinds of astrophysical observations, such as anomalies in the galactic rotation curves gravitational lensing effects in bullet cluster etc., have indicated the existence of DM in the Universe.
In the ITM, the extra scalar fields are protected by a discrete $\Z_2$-symmetry which ensures the stability of the lightest neutral particle.
We have verified that the mass of the neutral scalar particle ($H$) are slightly lighter than the mass of the charged particle ($H^\pm$) so that 
the contributions coming from co-annihilation between H and $H^\pm$ play a significant role in the relic density calculation. In the low mass
region, the co-annihilation rates are quite high so that the dark matter density is found to be much smaller than the right relic density $\Omega h^2=0.1198 \pm0.0026$ of the Universe.
We have obtained the relic density in the right ballpark for DM mass to be greater than $1.8$ TeV. 
In this context, we have shown how the presence of an additional hyperchargeless scalar triplet improves the stability of the Higgs potential.
In this study,
we have used state of the art next-to-next-to leading order (NNLO) for the SM calculations. We have used the SM Higgs scalar potential up to two-loop quantum corrections which is improved by three-loop renormalization groups of the SM couplings.
We have taken into account the contributions to the effective Higgs
potential of the new scalars at one-loop only. These contributions are improved by two-loop renormalization groups of the new parameters.
In this paper, we have explored the stability of the EW minimum of the new effective Higgs potential up to the Planck mass $\mpl$.
We have presented the new modified stability conditions for the metastable EW vacuum.
We have also shown various phase diagrams in various parameter spaces to show the explicit dependence of the EW (meta)stability on various parameters.
For the first time, we have identified new regions of parameter space that correspond to the stable and metastable EW vacuum, which also provides the relic density of the DM in the Universe as measured by the WMAP and Planck experiments.
In the present paper, we have also shown that the extra neutral scalar field $H$ can play the role of an inflation and can serve as a dark matter candidate. The scalar potential can explain inflation for large scalar field values. We have obtained the inflationary observables as observed by the experiments.

\vskip 20pt
\noindent{\bf Acknowledgements:}\\
The work of N.K. is supported by a fellowship from University Grants Commission. 
This work is partially supported by a grant from the Department of Science and Technology, India via Grant No. EMR/2014/001177. I would like to thank Subhendu Rakshit, Amitava Raychaudhuri, Amitava Datta, Subhendra Mohanty and Girish K. Chakravarty for useful discussions. 
\newpage
\appendix
\section{Two-loop beta functions for IT Model}
\label{App:BetaFunctionsTrip}
In this study, we use the SM RGEs up to three-loop which have been given in Refs.~\cite{Chetyrkin:2012rz,Zoller:2012cv,Chetyrkin:2013wya,Zoller:2013mra}. The triplet contributions ($\lambda_{2,3} $) are taken up to two-loop which have been generated using {\tt SARAH}~\cite{Staub:2013tta}.

In the HTM ($Y=0$), the RGEs of the couplings ($\chi_{i}=g_{1,2,3},\lambda_{1,2,3} $ and $Y_{l,u,d}$) and dimensionful mass parameters ($\mu_{1,2}$ and $\lambda_4$) are defined as 
\bea
\beta_{\chi_{i}}=\frac{\partial \chi_{i}}{\partial \ln \mu} =   \frac{1}{16 \pi^2}~\beta_{\chi_{i}}^{(1)}  +  \frac{1}{(16 \pi^2)^2}~\beta_{\chi_{i}}^{(2)}\, .\nn 
\eea

For $\mu>M_H$, the RGEs of the scalar quartic couplings $\lambda_{1,2,3}$ and the mass parameter $\lambda_4$ are given by
{\allowdisplaybreaks  \begin{align} 
\beta_{\lambda_1}^{(1)} & =  
+\frac{27}{200} g_{1}^{4} +\frac{9}{20} g_{1}^{2} g_{2}^{2} +\frac{9}{8} g_{2}^{4} -\frac{9}{5} g_{1}^{2} \lambda_1 -9 g_{2}^{2} \lambda_1 +24 \lambda_{1}^{2} +\frac{3}{2} \lambda_{3}^{2} +12 \lambda_1 \mbox{Tr}\Big({Y_d  Y_{d}^{\dagger}}\Big) +4 \lambda_1 \mbox{Tr}\Big({Y_l  Y_{l}^{\dagger}}\Big) \nonumber \\ 
 &+12 \lambda_1 \mbox{Tr}\Big({Y_u  Y_{u}^{\dagger}}\Big) -6 \mbox{Tr}\Big({Y_d  Y_{d}^{\dagger}  Y_d  Y_{d}^{\dagger}}\Big) -2 \mbox{Tr}\Big({Y_l  Y_{l}^{\dagger}  Y_l  Y_{l}^{\dagger}}\Big) -6 \mbox{Tr}\Big({Y_u  Y_{u}^{\dagger}  Y_u  Y_{u}^{\dagger}}\Big)\label{betal_11} \\
\beta_{\lambda_1}^{(2)} & =  
-\frac{3411}{2000} g_{1}^{6} -\frac{1677}{400} g_{1}^{4} g_{2}^{2} -\frac{317}{80} g_{1}^{2} g_{2}^{4} +\frac{277}{16} g_{2}^{6} +\frac{1887}{200} g_{1}^{4} \lambda_1 +\frac{117}{20} g_{1}^{2} g_{2}^{2} \lambda_1 -\frac{29}{8} g_{2}^{4} \lambda_1 +\frac{108}{5} g_{1}^{2} \lambda_{1}^{2} \nonumber \\ 
 &+108 g_{2}^{2} \lambda_{1}^{2} -312 \lambda_{1}^{3} +5 g_{2}^{4} \lambda_3 +12 g_{2}^{2} \lambda_{3}^{2} -15 \lambda_1 \lambda_{3}^{2} -2 \lambda_{3}^{3} \nonumber \\ 
 &+\frac{1}{20} \Big(-5 \Big(64 \lambda_1 \Big(-5 g_{3}^{2}  + 9 \lambda_1 \Big) -90 g_{2}^{2} \lambda_1  + 9 g_{2}^{4} \Big) + 9 g_{1}^{4}  + g_{1}^{2} \Big(50 \lambda_1  + 54 g_{2}^{2} \Big)\Big)\mbox{Tr}\Big({Y_d  Y_{d}^{\dagger}}\Big) \nonumber \\ 
 &-\frac{3}{20} \Big(15 g_{1}^{4}  -2 g_{1}^{2} \Big(11 g_{2}^{2}  + 25 \lambda_1 \Big) + 5 \Big(-10 g_{2}^{2} \lambda_1  + 64 \lambda_{1}^{2}  + g_{2}^{4}\Big)\Big)\mbox{Tr}\Big({Y_l  Y_{l}^{\dagger}}\Big) -\frac{171}{100} g_{1}^{4} \mbox{Tr}\Big({Y_u  Y_{u}^{\dagger}}\Big) \nonumber \\ 
 &+\frac{63}{10} g_{1}^{2} g_{2}^{2} \mbox{Tr}\Big({Y_u  Y_{u}^{\dagger}}\Big) -\frac{9}{4} g_{2}^{4} \mbox{Tr}\Big({Y_u  Y_{u}^{\dagger}}\Big) +\frac{17}{2} g_{1}^{2} \lambda_1 \mbox{Tr}\Big({Y_u  Y_{u}^{\dagger}}\Big) +\frac{45}{2} g_{2}^{2} \lambda_1 \mbox{Tr}\Big({Y_u  Y_{u}^{\dagger}}\Big) \nonumber \\ 
 &+80 g_{3}^{2} \lambda_1 \mbox{Tr}\Big({Y_u  Y_{u}^{\dagger}}\Big) -144 \lambda_{1}^{2} \mbox{Tr}\Big({Y_u  Y_{u}^{\dagger}}\Big) +\frac{4}{5} g_{1}^{2} \mbox{Tr}\Big({Y_d  Y_{d}^{\dagger}  Y_d  Y_{d}^{\dagger}}\Big) -32 g_{3}^{2} \mbox{Tr}\Big({Y_d  Y_{d}^{\dagger}  Y_d  Y_{d}^{\dagger}}\Big) \nonumber \\ 
 &-3 \lambda_1 \mbox{Tr}\Big({Y_d  Y_{d}^{\dagger}  Y_d  Y_{d}^{\dagger}}\Big) -42 \lambda_1 \mbox{Tr}\Big({Y_d  Y_{u}^{\dagger}  Y_u  Y_{d}^{\dagger}}\Big) -\frac{12}{5} g_{1}^{2} \mbox{Tr}\Big({Y_l  Y_{l}^{\dagger}  Y_l  Y_{l}^{\dagger}}\Big) - \lambda_1 \mbox{Tr}\Big({Y_l  Y_{l}^{\dagger}  Y_l  Y_{l}^{\dagger}}\Big) \nonumber \\ 
 &-\frac{8}{5} g_{1}^{2} \mbox{Tr}\Big({Y_u  Y_{u}^{\dagger}  Y_u  Y_{u}^{\dagger}}\Big) -32 g_{3}^{2} \mbox{Tr}\Big({Y_u  Y_{u}^{\dagger}  Y_u  Y_{u}^{\dagger}}\Big) -3 \lambda_1 \mbox{Tr}\Big({Y_u  Y_{u}^{\dagger}  Y_u  Y_{u}^{\dagger}}\Big) +30 \mbox{Tr}\Big({Y_d  Y_{d}^{\dagger}  Y_d  Y_{d}^{\dagger}  Y_d  Y_{d}^{\dagger}}\Big) \nonumber \\ 
 &+6 \mbox{Tr}\Big({Y_d  Y_{d}^{\dagger}  Y_d  Y_{u}^{\dagger}  Y_u  Y_{d}^{\dagger}}\Big) -12 \mbox{Tr}\Big({Y_d  Y_{u}^{\dagger}  Y_u  Y_{d}^{\dagger}  Y_d  Y_{d}^{\dagger}}\Big) -6 \mbox{Tr}\Big({Y_d  Y_{u}^{\dagger}  Y_u  Y_{u}^{\dagger}  Y_u  Y_{d}^{\dagger}}\Big) \nonumber \\ 
 &+10 \mbox{Tr}\Big({Y_l  Y_{l}^{\dagger}  Y_l  Y_{l}^{\dagger}  Y_l  Y_{l}^{\dagger}}\Big) +30 \mbox{Tr}\Big({Y_u  Y_{u}^{\dagger}  Y_u  Y_{u}^{\dagger}  Y_u  Y_{u}^{\dagger}}\Big) \label{betal_12}\\
\beta_{\lambda_2}^{(1)} & =  
2 \Big(11 \lambda_{2}^{2}  -12 g_{2}^{2} \lambda_2  + 3 g_{2}^{4}  + \lambda_{3}^{2}\Big)\\ 
\beta_{\lambda_2}^{(2)} & =  
-\frac{272}{3} g_{2}^{6} +\frac{94}{3} g_{2}^{4} \lambda_2 +160 g_{2}^{2} \lambda_{2}^{2} -244 \lambda_{2}^{3} +10 g_{2}^{4} \lambda_3 +\frac{12}{5} g_{1}^{2} \lambda_{3}^{2} +12 g_{2}^{2} \lambda_{3}^{2} -20 \lambda_2 \lambda_{3}^{2} -8 \lambda_{3}^{3} \nonumber \\ 
 &-12 \lambda_{3}^{2} \mbox{Tr}\Big({Y_d  Y_{d}^{\dagger}}\Big) -4 \lambda_{3}^{2} \mbox{Tr}\Big({Y_l  Y_{l}^{\dagger}}\Big) -12 \lambda_{3}^{2} \mbox{Tr}\Big({Y_u  Y_{u}^{\dagger}}\Big) \\
\beta_{\lambda_3}^{(1)} & =  
+3 g_{2}^{4} -\frac{9}{10} g_{1}^{2} \lambda_3 -\frac{33}{2} g_{2}^{2} \lambda_3 +12 \lambda_1 \lambda_3 +10 \lambda_2 \lambda_3 +4 \lambda_{3}^{2} +6 \lambda_3 \mbox{Tr}\Big({Y_d  Y_{d}^{\dagger}}\Big) +2 \lambda_3 \mbox{Tr}\Big({Y_l  Y_{l}^{\dagger}}\Big) \nonumber \\ 
 &+6 \lambda_3 \mbox{Tr}\Big({Y_u  Y_{u}^{\dagger}}\Big) \\
\beta_{\lambda_3}^{(2)} & =  
-\frac{9}{4} g_{1}^{2} g_{2}^{4} +\frac{329}{12} g_{2}^{6} +30 g_{2}^{4} \lambda_1 +20 g_{2}^{4} \lambda_2 +\frac{1671}{400} g_{1}^{4} \lambda_3 +\frac{9}{8} g_{1}^{2} g_{2}^{2} \lambda_3 -\frac{607}{48} g_{2}^{4} \lambda_3 +\frac{72}{5} g_{1}^{2} \lambda_1 \lambda_3 \nonumber \\ 
 &+72 g_{2}^{2} \lambda_1 \lambda_3 -60 \lambda_{1}^{2} \lambda_3 +88 g_{2}^{2} \lambda_2 \lambda_3 -34 \lambda_{2}^{2} \lambda_3 +\frac{3}{5} g_{1}^{2} \lambda_{3}^{2} +11 g_{2}^{2} \lambda_{3}^{2} -72 \lambda_1 \lambda_{3}^{2} -52 \lambda_2 \lambda_{3}^{2} \nonumber \\ 
 &-\frac{23}{2} \lambda_{3}^{3} +\Big(-12 \lambda_{3}^{2}  -3 g_{2}^{4}  + 40 g_{3}^{2} \lambda_3  -72 \lambda_1 \lambda_3  + \frac{45}{4} g_{2}^{2} \lambda_3  + \frac{5}{4} g_{1}^{2} \lambda_3 \Big)\mbox{Tr}\Big({Y_d  Y_{d}^{\dagger}}\Big) \nonumber \\ 
 &-\frac{1}{4} \Big(4 g_{2}^{4}  + \lambda_3 \Big(16 \lambda_3  + 96 \lambda_1 -15 g_{2}^{2}-15 g_{1}^{2} \Big)\Big)\mbox{Tr}\Big({Y_l  Y_{l}^{\dagger}}\Big) -3 g_{2}^{4} \mbox{Tr}\Big({Y_u  Y_{u}^{\dagger}}\Big) +\frac{17}{4} g_{1}^{2} \lambda_3 \mbox{Tr}\Big({Y_u  Y_{u}^{\dagger}}\Big) \nonumber \\ 
 &+\frac{45}{4} g_{2}^{2} \lambda_3 \mbox{Tr}\Big({Y_u  Y_{u}^{\dagger}}\Big) +40 g_{3}^{2} \lambda_3 \mbox{Tr}\Big({Y_u  Y_{u}^{\dagger}}\Big) -72 \lambda_1 \lambda_3 \mbox{Tr}\Big({Y_u  Y_{u}^{\dagger}}\Big) -12 \lambda_{3}^{2} \mbox{Tr}\Big({Y_u  Y_{u}^{\dagger}}\Big) \nonumber \\ 
 &-\frac{27}{2} \lambda_3 \mbox{Tr}\Big({Y_d  Y_{d}^{\dagger}  Y_d  Y_{d}^{\dagger}}\Big) -21 \lambda_3 \mbox{Tr}\Big({Y_d  Y_{u}^{\dagger}  Y_u  Y_{d}^{\dagger}}\Big) -\frac{9}{2} \lambda_3 \mbox{Tr}\Big({Y_l  Y_{l}^{\dagger}  Y_l  Y_{l}^{\dagger}}\Big) -\frac{27}{2} \lambda_3 \mbox{Tr}\Big({Y_u  Y_{u}^{\dagger}  Y_u  Y_{u}^{\dagger}}\Big)
\end{align}} 
{\allowdisplaybreaks  \begin{align}
\beta_{\lambda_4}^{(1)} & =  
2 \lambda_4 \mbox{Tr}\Big({Y_l  Y_{l}^{\dagger}}\Big)  + 4 \lambda_1 \lambda_4  + 4 \lambda_3 \lambda_4  + 6 \lambda_4 \mbox{Tr}\Big({Y_d  Y_{d}^{\dagger}}\Big)  + 6 \lambda_4 \mbox{Tr}\Big({Y_u  Y_{u}^{\dagger}}\Big)  -\frac{21}{2} g_{2}^{2} \lambda_4  -\frac{9}{10} g_{1}^{2} \lambda_4 \\ 
\beta_{\lambda_4}^{(2)} & =  
+\frac{1311}{400} g_{1}^{4} \lambda_4 +\frac{141}{40} g_{1}^{2} g_{2}^{2} \lambda_4 -\frac{1343}{48} g_{2}^{4} \lambda_4 +\frac{24}{5} g_{1}^{2} \lambda_1 \lambda_4 -28 \lambda_{1}^{2} \lambda_4 +5 \lambda_{2}^{2} \lambda_4 +\frac{3}{5} g_{1}^{2} \lambda_3 \lambda_4 \nonumber \\ 
 &+23 g_{2}^{2} \lambda_3 \lambda_4 -40 \lambda_1 \lambda_3 \lambda_4 -20 \lambda_2 \lambda_3 \lambda_4 -\frac{17}{2} \lambda_{3}^{2} \lambda_4 +\frac{\lambda_4}{4} \Big(160 g_{3}^{2}  + 45 g_{2}^{2}  -48 \lambda_3  + 5 g_{1}^{2} \nonumber \\ 
 & -96 \lambda_1 \Big) \mbox{Tr}\Big({Y_d  Y_{d}^{\dagger}}\Big) +\frac{1}{4} \Big(15 g_{1}^{2}  + 15 g_{2}^{2}  -16 \Big(2 \lambda_1  + \lambda_3\Big)\Big)\lambda_4 \mbox{Tr}\Big({Y_l  Y_{l}^{\dagger}}\Big) \nonumber \\ 
 &+\frac{17}{4} g_{1}^{2} \lambda_4 \mbox{Tr}\Big({Y_u  Y_{u}^{\dagger}}\Big) +\frac{45}{4} g_{2}^{2} \lambda_4 \mbox{Tr}\Big({Y_u  Y_{u}^{\dagger}}\Big) +40 g_{3}^{2} \lambda_4 \mbox{Tr}\Big({Y_u  Y_{u}^{\dagger}}\Big) -24 \lambda_1 \lambda_4 \mbox{Tr}\Big({Y_u  Y_{u}^{\dagger}}\Big) \nonumber \\ 
 &-12 \lambda_3 \lambda_4 \mbox{Tr}\Big({Y_u  Y_{u}^{\dagger}}\Big) -\frac{27}{2} \lambda_4 \mbox{Tr}\Big({Y_d  Y_{d}^{\dagger}  Y_d  Y_{d}^{\dagger}}\Big) +27 \lambda_4 \mbox{Tr}\Big({Y_d  Y_{u}^{\dagger}  Y_u  Y_{d}^{\dagger}}\Big) -\frac{9}{2} \lambda_4 \mbox{Tr}\Big({Y_l  Y_{l}^{\dagger}  Y_l  Y_{l}^{\dagger}}\Big) \nonumber \\ 
 &-\frac{27}{2} \lambda_4 \mbox{Tr}\Big({Y_u  Y_{u}^{\dagger}  Y_u  Y_{u}^{\dagger}}\Big).
\end{align}} 

For $\mu<M_H$, $ \beta_{\lambda_1} = \beta_{\lambda_1} (\lambda_{2,3}=0)$ and $\beta_{\lambda_{2,3,4}} =0$, where, $Y_u=y_u,~y_c,~y_t$ are the Yukawa couplings of up-,charm- and top-quark, $Y_d=y_d,~y_s,~y_b$ for down-, strange- and bottom-quark. $Y_l$ represents the Yukawa couplings for the charged leptons. In our work, we have included the contribution only from top-quark. Since, the other Yukawa couplings are very small, they do not alter our result. We have also taken into account the contributions to the beta functions of the gauge couplings $g_{1,2,3}$ of the new physics. The importance of mass parameters $\mu_{1,2}$ and $\lambda_4$ are found to be negligible in the stability analysis.


\end{document}